\documentclass[aps,prd,superscriptaddress,showpacs,twocolumn]{revtex4}

\usepackage{amsfonts}
\usepackage{amsmath}
\usepackage{graphicx}
\usepackage{color}
\usepackage{amssymb}

\usepackage{soul}

\usepackage[normalem]{ulem}

\begin{document}

\title{Modification of transition radiation by three-dimensional topological insulators}

\author{O. J. Franca}
\email{francamentesantiago@ciencias.unam.mx}
\affiliation{Institut f\"ur Physik, Universit\"at Kassel, Heinrich-Plett-Stra\ss e 40, 34132 Kassel, Germany}

\author{Stefan Yoshi Buhmann}
\email{stefan.buhmann@uni-kassel.de}
\affiliation{Institut f\"ur Physik, Universit\"at Kassel, Heinrich-Plett-Stra\ss e 40, 34132 Kassel, Germany}

\begin{abstract}
We study how transition radiation is modified by the presence of a generic magnetoelectric medium with a special focus on topological insulators. To this end, we use the Green's function for the electromagnetic field in presence of a plane interface between two topological insulators with different topological parameters, permittivities, and permeabilities. We employ the far-field approximation together with the steepest descent method to obtain approximate analytical expressions for the electromagnetic field. Through this method, we find that the electric field is a superposition of spherical waves and lateral waves. Contributions of both kinds can be attributed to a purely topological origin. After computing the angular distribution of the radiation, we find that in a region far from the interface the main contribution to the radiation comes from the spherical waves. We present typical radiation patterns for the topological insulator TlBiSe$_2$ and the magnetoelectric TbPO$_4$. In the ultra-relativistic case, the additional contributions from the magnetoelectric coupling appreciably enhance the global maximum of the angular distribution. We also present an analytic expression for the frequency distribution of the radiation for this case. We find that in the limit where the permittivities are equal, there still exists transition radiation of the order of the square of the topological parameter with a pure topological origin.

\end{abstract}


\maketitle

\section{Introduction} \label{INTRO}
Once the Vavilov-\v{C}erenkov radiation was discovered \cite{Cherenkov, Vavilov} and understood \cite{FT}, the paradigm that a particle with constant velocity traveling through a medium cannot radiate was broken down. As confirmation of the paradigm shift a few years later, the theoretical development of transition radiation \cite{Frank-Ginzburg} and its experimental verification \cite{Goldsmith-Jelley} opened a gate to new research in electromagnetic radiation. Transition radiation is a phenomenon that occurs when a charged particle with constant velocity crosses an interface between two media with different permittivities and permeabilities. The original work of Ginzburg and Frank \cite{Frank-Ginzburg} showed the existence of backwards radiation emitted relative to the particle motion, which constituted an important distinction with the Vavilov-\v{C}erenkov radiation that is emitted forwards in right-handed homogeneous media. Additionally, it was shown that the frequency of this backwards radiation was mainly in the visible range of the electromagnetic spectrum and that the radiation intensity was logarithmically proportional to the Lorentz factor of the particle. Interestingly, if the media is left-handed the Vavilov-\v{C}erenkov radiation can be emitted backwards too. This exotic electromagnetic phenomenon was predicted by Veselago \cite{Veselago} and subsequently studied in Refs. \cite{Pendry, Tao, Duan, Alekhina}.

Despite the first measurement of transition radiation in the optical region by Goldsmith and Jelley \cite{Goldsmith-Jelley} the interest devoted to this subject was not increased, because the low intensity of the radiation seemed to prohibit its use for the detection and identification of individual particles. The situation changed when Garibian \cite{Garibian} demonstrated that transition radiation by an ultrarelativistic particle will be emitted in the x-ray regime. Shortly after, Garibian \cite{Garibian 2} studied a particle emitting transition radiation by passing through a layer of matter of finite thickness and he found that the energy lost was directly proportional to the Lorentz factor of the particle in the ultrarelativistic regime. This theoretical discovery of x-ray transition radiation paved the way towards using transition radiation in high-energy physics as a tool for particle identification at high momenta \cite{Cherry et al}. This was achieved by implementing and designing transition radiation detectors \cite{Dolgoshein, Andronic}, which have been used and are currently being used or planned in a wide range of accelerator-based experiments, such as UA2 \cite{Ansari et al}, ZEUS \cite{Appuhn et al}, NA31 \cite{Barr et al}, PHENIX \cite{OBrien et al 1, OBrien et al 2}, HELIOS \cite{Dolgoshein}, D$\emptyset$ \cite{Detoeuf et al,Piekarz}, kTeV \cite{Graham}, H1 \cite{Graessler,Beck}, WA89 \cite{Brueckner}, NOMAD \cite{Bassompierre}, HERMES \cite{Ackerstaff}, HERA-B \cite{Saveliev}, ATLAS \cite{Aad et al}, ALICE \cite{Aamodt et al}, CBM \cite{Andronic CBM} and in astro-particle and cosmic-ray experiments: WIZARD \cite{Belotti}, HEAT \cite{Barwick et al}, MACRO \cite{Ambrosio}, AMS \cite{Doetinchem}, PAMELA \cite{Ambriola}, ACCESS \cite{Case et al}.

Besides the applications for high-energy physics, transition radiation has also been studied in new kinds of materials such as chiral matter \cite{Galyamin} and even in photonic topological crystals \cite{Yu et al}. Other new interesting type of materials are topological insulators (TIs), to which this paper is devoted. Three-dimensional TIs  are a novel state of matter exhibiting insulating bulk and conducting surface states protected by time-reversal symmetry \cite{Hasan,Qi Review}. Their topological behavior was predicted in Refs. \cite{TIs Prediction 1,TIs Prediction 2} where two types have been distinguished: strong three-dimensional TIs exhibit an odd number of surface states on each surface, whereas the weak ones have an even number of them on all but a single surface \cite{TIs Prediction 1, Morgenstern review}. The experimental discovery of the first strong three-dimensional TI in Bi$_{1-x}$Sb$_x$ was confirmed in the work \cite{Hsieh}. This led to the detection of a huge variety of TIs as reported in \cite{TIs variety} and the more recently discovered new class of topological insulators, called axion insulators \cite{AXIs}. They have the same bulk as TIs, with gapped bulk and surface states, where the topological properties are protected by inversion, instead of time reversal symmetry. They can arise up in magnetically doped-TI heterostructures  with magnetization pointing inwards and outwards from the top and bottom interfaces of the TI \cite{AXIs}. When time-reversal symmetry is broken at the interface between a 3D strong TI and a regular insulator, either by the application of a magnetic coating and/or by doping the TI with transition metal elements, the opening of the gap in the surface states induces several exotic phenomena which can be tested experimentally. Among them, we find the quantum anomalous Hall effect \cite{QAH,Mogi}, the quantized magneto-optical effect \cite{Wu, Dziom}, the topological magnetoelectric effect \cite{Hasan, Qi PRB, Okada} and even radiative effects as the reversed Vavilov-\v{C}erenkov radiation \cite{OJF-LFU-ORT}, which was recently theoretically predicted. Also the image magnetic monopole effect has been analyzed in Ref. \cite{Qi Science} but its direct experimental verification still remains an open challenge. These effects can be also produced in axion insulators.

The aim of the present work is to study how transition radiation is modified by strong three-dimensional TIs by combining the analysis of usual transition radiation given in the references \cite{Frank-Ginzburg, Frank Lecture, Ginzburg} with the Green's function method that we will present here. This paper is organized as follows. In Sec. \ref{EFT}, we describe the modified Maxwell equations which describe the electromagnetic response of three-dimensional TIs from an effective field theory point of view. Also in this section, we review the Green's function method to obtain the time-dependent electromagnetic field arising from arbitrary sources in the presence of two semi-infinite magnetoelectric media with different values of the topological parameter, permittivities and permeabilities separated by a planar interface that encodes the manifestation of the topological magneteoelectric effect. Section \ref{CHARGE} is devoted to analyzing the electromagnetic field associated with transition radiation. To this end, we consider a charged particle moving with constant velocity $v$ in the direction perpendicular to the interface between the two magnetoelectrics. Our formalism yields analytic results for the physical quantities involved. The far-field expressions for the electric  and magnetic induction fields are calculated. In Sec. \ref{ANGULAR}, we present the angular distribution of the transition radiation  for two cases: (i) when the observer is located at the upper magnetoelectric medium and the incident particle parts from this medium towards the lower one, and (ii) when the trajectory of the particle is reversed and the observer is at the lower magnetoelectric medium. Here we present radiation patterns to illustrate our results and the differences between this transition radiation and the usual one. We analyze different limiting cases, finding that when the permittivities and permeabilities are equal, there still exists transition radiation of the order of the square of the topological parameter. Section \ref{ULTRA} is devoted to obtain an analytical expression for the frequency distribution of transition radiation in the ultrarelativistic case. We focus on two cases: when the particle moves from vacuum into a nonmagnetical strong three-dimensional TI, referred as vacuum-to-TI case, and the swapped case dubbed as TI-to-vacuum case. Section \ref{CONCLUSIONS} comprises a concluding summary of our results.

\section{Modified Maxwell Equations} \label{EFT}

As in all studies regarding the electromagnetic response of a certain system, our starting point will be the Maxwell equations and constitutive relations of the involved material. For a three-dimensional TI these are \cite{Qi PRB,Chang-Yang,Obukhov-Hehl}
\begin{eqnarray}
\nabla\cdot\mathbf{B}\left(\mathbf{r};\omega\right) &=& 0 \;, \label{Gauss B}\\
\nabla\times\mathbf{E}\left(\mathbf{r};\omega\right) &=& \mathrm{i}\omega \mathbf{B}\left(\mathbf{r};\omega\right) \;, \label{Faraday}\\
\nabla\cdot\mathbf{D}\left(\mathbf{r};\omega\right) &=& \varrho_E \left(\mathbf{r};\omega\right) \;, \label{Gauss D}\\
\nabla\times\mathbf{H}\left(\mathbf{r};\omega\right) + \mathrm{i} \omega \mathbf{D}\left(\mathbf{r};\omega\right) &=& \mathbf{j}_E\left(\mathbf{r};\omega\right) \;, \label{Ampere-Maxwell}
\end{eqnarray}
where $\varrho_E(\mathbf{r};\omega)$  and  $\mathbf{j}_E (\mathbf{r};\omega)$ denote the source term for electromagnetic waves generated by external charges and currents, respectively, $\mathbf{E}\left(\mathbf{r};\omega\right)$ stands for the electric field and $\mathbf{B}\left(\mathbf{r};\omega\right)$ is the induction field. For these materials, both fields are connected with the displacement field $\mathbf{D}\left(\mathbf{r};\omega\right)$ and the magnetic field $\mathbf{H}\left(\mathbf{r};\omega\right)$ via
\begin{eqnarray}
\mathbf{D}\left(\mathbf{r};\omega\right) &=& \varepsilon_0\varepsilon\left(\mathbf{r};\omega\right) \mathbf{E}\left(\mathbf{r};\omega\right) + \frac{ \alpha \Theta\left(\mathbf{r};\omega\right) }{ \pi\mu_0 c }  \mathbf{B}\left(\mathbf{r};\omega\right) \nonumber\\
&& + \mathbf{P}_N\left(\mathbf{r};\omega\right) \;, \label{Constitutive 1}
\end{eqnarray}
\begin{eqnarray}
\mathbf{H}\left(\mathbf{r};\omega\right) &=& \frac{ \mathbf{B}\left(\mathbf{r};\omega\right) }{ \mu_0\mu\left(\mathbf{r};\omega\right) } -  \frac{ \alpha \Theta\left(\mathbf{r};\omega\right) }{ \pi\mu_0 c }  \mathbf{E}\left(\mathbf{r};\omega\right) \nonumber\\
&& - \mathbf{M}_N\left(\mathbf{r};\omega\right) \;, \label{Constitutive 2}
\end{eqnarray}
where $\alpha$ is the fine structure constant and $\varepsilon\left(\mathbf{r};\omega\right)$, $\mu\left(\mathbf{r};\omega\right)$, and $\Theta\left(\mathbf{r};\omega\right)$ are the dielectric permittivity, magnetic permeability, and axion coupling, respectively. Here the $\mathbf{P}_N\left(\mathbf{r};\omega\right)$ and $\mathbf{M}_N\left(\mathbf{r};\omega\right)$ terms are the noise polarization and magnetization, respectively. These are Langevin noise terms that model absorption within the material \cite{Scheel-Buhmann}.  

The axion coupling $\Theta\left(\mathbf{r};\omega\right)$ takes even multiples of $\pi$ in weak three-dimensional TIs and odd multiples of $\pi$ in strong three-dimensional TIs, with the magnitude and sign of the multiple given by the strength and direction of the time-symmetry-breaking perturbation. It is worth noting that the modified Maxwell Eqs. (\ref{Gauss B})-(\ref{Ampere-Maxwell}) can be derived from the Lagrangian density $\mathcal{L}=\mathcal{L}_0 + \mathcal{L}_{\Theta}$, where $\mathcal{L}_0$ is the usual Maxwell Lagrangian density and $\mathcal{L}_{\Theta}$ is given by
\begin{equation}
\mathcal{L}_{\Theta} =  \frac{ \alpha }{ 4\pi^2 } \frac{ \Theta\left(\mathbf{r};\omega\right) }{ \mu_0 c } \mathbf{E}\left(\mathbf{r};\omega\right) \cdot \mathbf{B}\left(\mathbf{r};\omega\right) \;.
\end{equation}
In the context of condensed matter, $\Theta$ is known as the (scalar) magnetoelectric polarizability \cite{Essin}, but in field theory, this quantity is called the axion field, which generates axion electrodynamics and whose Lagrangian is just the above one \cite{Wilczek}. In this way, for three-dimensional TIs, this additional $\Theta$ term induces field-dependent effective charge and current densities which model the topological magnetoelectric effect \cite{Hasan,Qi PRB}. From a broader perspective, bi-isotropic materials couple the electric and magnetic field to each other by means of magnetoelectric parameters $\chi$ and $\kappa$ characterizing nonreciprocity and chirality respectively. We can regard the constitutive relations (\ref{BC1}) and (\ref{BC2}) as those for a bi-isotropic material plate $\mathbf{D}=\varepsilon_0\varepsilon\mathbf{E}+(\chi-\mathrm{i} \kappa)\sqrt{\varepsilon_0\mu_0}\,\mathbf{H}$ and $\mathbf{B}=\mu_0\mu\mathbf{H}+(\chi+\mathrm{i} \kappa)\sqrt{\varepsilon_0\mu_0}\,\mathbf{E}$ yielding the identifications $\chi=\alpha\mu\Theta\left(\mathbf{r};\omega\right)/\pi$ and $\kappa=0$. In this framework, the axion coupling endows the nonreciprocal feature to the TIs which breaks time-reversal symmetry as mentioned in Introduction. Moreover, TIs are classified as Tellegen media because they do not present any chirality \cite{Jiang}. 

Using the above constitutive relations (\ref{Constitutive 1}) and (\ref{Constitutive 2}), one can show that the frequency components of the electric field obey the inhomogeneous Helmholtz equation 
\begin{eqnarray}
&&\nabla \times \frac{ 1 }{ \mu(\mathbf{r};\omega) } \nabla \times \mathbf{E}(\mathbf{r};\omega) - \frac{ \omega^2 }{ c^2 }\varepsilon(\mathbf{r};\omega) \mathbf{E}(\mathbf{r};\omega) \nonumber\\
&& - \mathrm{i} \frac{ \omega \alpha }{ \pi c } \nabla \Theta(\mathbf{r};\omega) \times \mathbf{E}(\mathbf{r};\omega) = \mathrm{i} \omega \mu_0 \mathbf{j}(\mathbf{r};\omega)\;,  \label{inhomo Helmholtz}
\end{eqnarray}
where $\mathbf{j}(\mathbf{r};\omega)=\mathbf{j}_E (\mathbf{r};\omega) + \mathbf{j}_N (\mathbf{r};\omega)$ is the total current density,  and $\mathbf{j}_N (\mathbf{r};\omega)=-\mathrm{i}\omega \mathbf{P}_N (\mathbf{r};\omega) + \nabla\times\mathbf{M}_N(\mathbf{r};\omega)$ is the source term for electromagnetic waves generated by noise fluctuations within the material. If the axion coupling is homogeneous, $\Theta(\mathbf{r};\omega)=\Theta(\omega)$, then the last term on the left-hand side vanishes and one finds that the propagation of the electric field is the same as in a conventional magnetoelectric. As a result, electromagnetic waves propagating within a homogeneous three-dimensional TI retain their usual properties. Thus, the effects of the axion coupling are only felt when the axion coupling depends on spatial coordinates, as it happens at the interface of two materials having different constant values of $\Theta$, for example. When $\Theta$ is restricted to a nondynamical field, the theory is usually referred as $\Theta$-Electrodynamics \cite{OJF-LFU-ORT, AMR-LFU-MC 1, AMR-LFU-MC 2}, which has the necessary topological features to model the electromagnetic behavior of three-dimensional TIs \cite{Qi PRB} and also naturally existing magnetoelectric media \cite{Hehl,Landau-Lifshitz} such as Cr$_2$O$_3$ with $\Theta\simeq\pi/36$ \cite{Wu}.

\begin{figure}[tbp]
\centering 
\includegraphics[width=8CM]{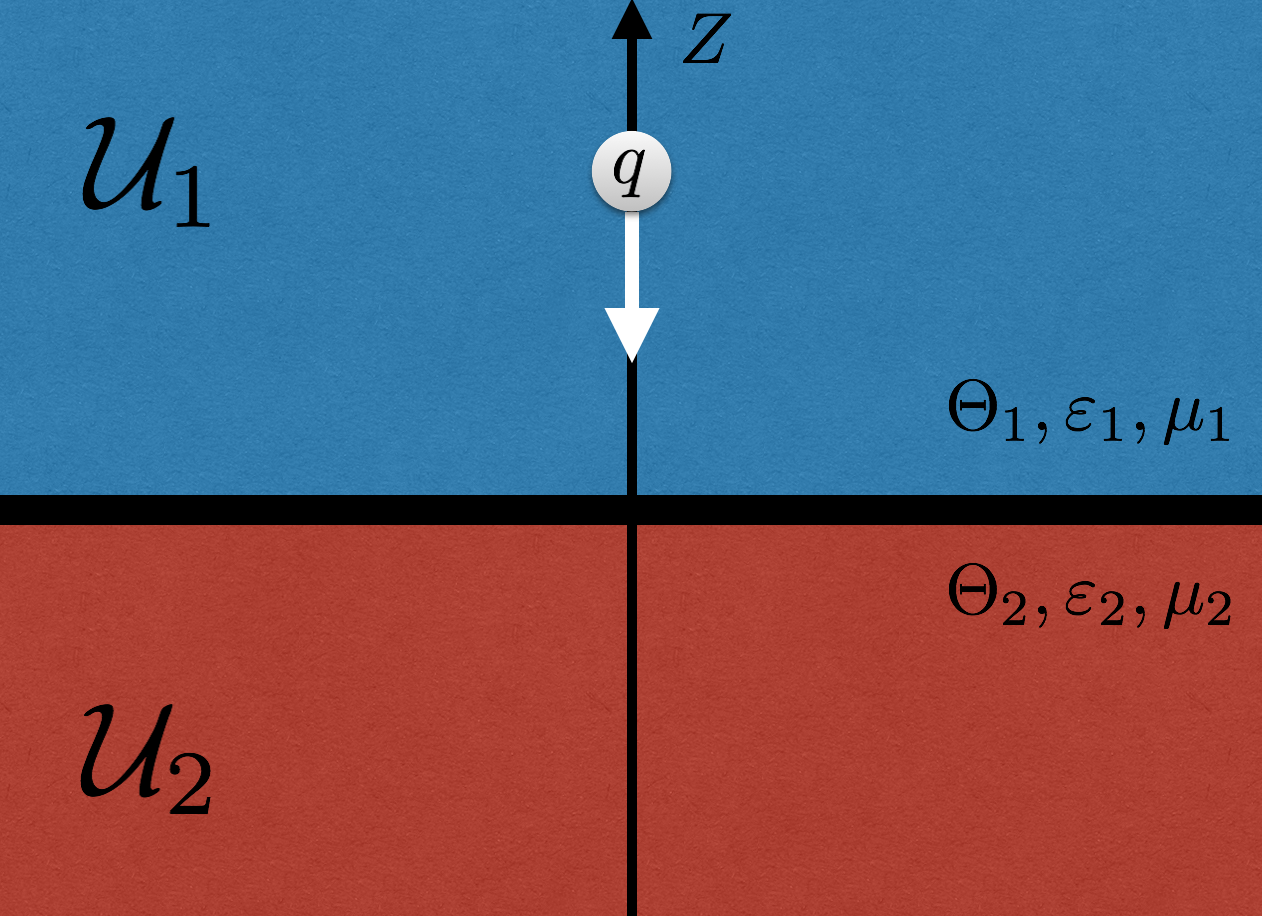} 
\caption{ Two semi-infinite magnetoelectric media with different magnetoelectric polarizabilities $\Theta_1$ and $\Theta_2$, having different permittivities $\varepsilon_1$ and $\varepsilon_2$, different permeabilities $\mu_1$ and $\mu_2$, and separated by the planar interface. }
\label{REGIONS}
\end{figure}

In order to solve Eq. (\ref{inhomo Helmholtz}) that dictates the dynamics for the electric field, we will employ the Green's function method, which is the solution to the wave equation (\ref{inhomo Helmholtz}) for a single frequency point source. Its advantage lies in the ability to compute $\mathbf{E}(\mathbf{r};\omega)$ at any point from an arbitrary distribution of current sources, once one has at hand the Green's function of the desired configuration. In our case, let us consider two semi-infinite magnetoelectric media separated by a planar interface located at $z=0$, filling the regions $\mathcal{U}_1$ and $\mathcal{U}_2$ of the space, as shown in Fig. \ref{REGIONS}. Additionally, we assume that the axion coupling $\Theta$ is piecewise constant taking the values $\Theta=\Theta_1$ in the region $\mathcal{U}_1$ and $\Theta=\Theta_2$ in the region $\mathcal{U}_2$. This is expressed as 
\begin{equation}\label{THETA}
\Theta(z) = \Theta_1 H(z) + \Theta_2 H(-z) \;,
\end{equation}
where $H(z)$ is the Heaviside function.

Fortunately, the Green's function $\mathbb{G}(\mathbf{r},\mathbf{r}';\omega)$ for this configuration was already deduced in Ref. \cite{Crosse-Fuchs-Buhmann}, which in particular connects the distribution of current sources to the electric field via 
\begin{equation} \label{E GF j}
\mathbf{E}(\mathbf{r};\omega) = \mathrm{i}\omega\mu_0 \int d^{3}\mathbf{r}\, \mathbb{G}(\mathbf{r},\mathbf{r}';\omega) \cdot\mathbf{j}(\mathbf{r}';\omega)\;.
\end{equation}

Two important features of the Green's function should be remarked. The first one is related to its reflective and transmissive behaviors. As can be seen from Fig. \ref{REGIONS}, for a field point at $z>0$ and if the source is in region $\mathcal{U}_1$ ($z'>0$) there are two contributions, one from direct propagation from the source to the field point, which is given by the free space Green's function $\mathbb{G}^{(0)}(\mathbf{r},\mathbf{r}';\omega)$, and one from reflections from the surface, which is given by the reflective part of the Green's function $\mathbb{G}^{(1)}(\mathbf{r},\mathbf{r}';\omega)$. For $z<0$, the only contribution is from the transmission at the surface, which is given by the transmissive part of the Green's function $\mathbb{G}^{(1)}(\mathbf{r},\mathbf{r}';\omega)$. Thus we can split the Green's function into two main parts
\begin{equation} \label{GF split}
\mathbb{G}(\mathbf{r},\mathbf{r}';\omega) = \left\{
\begin{array}{cc}
\mathbb{G}^{(0)}(\mathbf{r},\mathbf{r}';\omega) + \mathbb{G}^{(1)}(\mathbf{r},\mathbf{r}';\omega) \;, & z>0 \;,     \\
\mathbb{G}^{(1)}(\mathbf{r},\mathbf{r}';\omega) \;, & z<0 \;,    
\end{array}
\right.
\end{equation}
each of which can be computed separately. All the components of this Green's function can be found in Ref. \cite{Crosse-Fuchs-Buhmann}. The second feature of this Green's function comes from the nature of three-dimensional TIs or magnetoelectric media, which it describes. As it is well-known TIs break time-reversal-symmetry meaning that they can be classified as noncreciprocal media \cite{Fuchs-Crosse-Buhmann}. A nonreciprocal medium violates time-reversal symmetry and, hence, the Lorentz reciprocity principle for the Green's tensor \cite{Buhmann 1}
\begin{equation}
\mathbb{G}(\mathbf{r},\mathbf{r}';\omega) \neq \mathbb{G}^{T}(\mathbf{r}',\mathbf{r};\omega) \;.
\end{equation}
The breakdown of this principle has the important consequence that the symmetry relations \cite{Buhmann 1}
\begin{equation}\label{Usual Fresnel}
r_{\sigma}^{12} = -r_{\sigma}^{21} \;\; , \;\; \frac{ \mu_2(\omega) }{ k_{z,2} } t_{\sigma}^{21} = \frac{ \mu_1(\omega) }{ k_{z,1} } t_{\sigma}^{12} \;, (\sigma=\text{TM}, \text{TE})\;,
\end{equation}
for the standard Fresnel coefficients are no longer valid for our configuration.  Here $r_{\sigma}^{12}$ stands for the reflexion Fresnel coefficient with polarization $\sigma$, whose superscripts denotes the medium 1 as an upper layer and the medium 2 as the lower one or vice versa. Analogously, this notation applies for the transmission Fresnel coefficient $t_{\sigma}^{12}$. We denote by $k_{z,j}=\sqrt{k^2_j-\mathbf{k}_\parallel^2}$ the $z$-component of the wave vector $\mathbf{k}_{j\pm}=(k_{x}, k_{y}, \pm k_{z,j})$ with wavenumber $k_j=\sqrt{ \varepsilon_j(\omega) \mu_j(\omega) }\,\omega/c$ for $j=1,2$. Henceforth, every subscript on a physical quantity will denote that it is related to the medium 1 or medium 2.

In addition to Eq. (\ref{inhomo Helmholtz}), it is important to consider the boundary conditions that the electric and magnetic field must satisfy. Assuming that the time derivatives of the fields are finite in the vicinity of the interface at $z=0$ as well as the noise terms, the modified Maxwell equations (\ref{Gauss B})-(\ref{Ampere-Maxwell}) and the constitutive relations (\ref{Constitutive 1}) and (\ref{Constitutive 2}) yield the following boundary conditions:
\begin{eqnarray}
&& \left[ \varepsilon_0\varepsilon \mathbf{E}_{z} \right]_{z=0^{-}}^{z=0^{+}} = \frac{ \Delta \mathbf{B}_{z}|_{z=0} }{ c\mu_0\mu_1\mu_2 }\;,\; \left[\mathbf{B}_{z}\right]_{z=0^{-}}^{z=0^{+}} = \mathbf{ 0 } \;, \label{BC1}\\
&& \left[ \frac{ \mathbf{B}_{\parallel} }{ \mu_0\mu } \right]_{z=0^{-}}^{z=0^{+}} = -\frac{ \Delta \mathbf{E}_{\parallel}|_{z=0} }{ c\mu_0\mu_1\mu_2 } \;,\; \left[\mathbf{E}_{\parallel}\right]_{z=0^{-}}^{z=0^{+}} = \mathbf{ 0 } \;, \label{BC2}
\end{eqnarray}
for vanishing external sources at the interface, where  
\begin{equation} \label{DELTA}
\Delta=\alpha\mu_1\mu_2(\Theta_2-\Theta_1)/\pi \;.
\end{equation}
The notation is $\left[ \mathbf{V}\right]_{z=0^{-}}^{z=0^{+}}=\mathbf{V}(z=0^{+})-\mathbf{V}(z=0^{-})$, $\mathbf{V}\big|_{z=0}=\mathbf{V}(z=0)$, where $z=0^{\pm }$ indicates the limits $z=0\pm a $, with $a$ a positive real number such that $\; a\rightarrow 0$, and $\mathbf{V}$ an arbitrary vector field, respectively.

In the case of a strong three-dimensional TI located in region $\mathcal{U}_2$  of Fig. \ref{REGIONS} ($\Theta_1=\pi$) in front of a regular insulator ($\Theta_2=0$) in region $\mathcal{U}_2$,  we have 
\begin{equation}\label{DELTA_1}
\Delta=\alpha(2m+1), 
\end{equation}
where $m$ is an integer depending on the details of the time-reversal-symmetry breaking at the interface.

%
\section{Electromagnetic field of the Transition Radiation} \label{CHARGE}
Let us now consider a particle with charge $q$ and constant velocity $v\mathbf{e}_z$, perpendicular to the interface defined by the $xy$ plane ($z=0$) where $\mathbf{e}_{z}=(0,0,1)$, as shown in Fig. \ref{REGIONS}. The external charge and current densities in frequency space are
\begin{eqnarray}
\varrho_E(\mathbf{r}';\omega) &=& -\frac{q}{v}\delta(x')\delta(y')e^{-i\omega\frac{z'}{v}} \;, \\
\mathbf{j}_E(\mathbf{r}';\omega) &=& q\delta(x')\delta(y')e^{-i\omega\frac{z'}{v}}\mathbf{e}_z\;,  \label{rho y j}
\end{eqnarray}
where $v$ is assumed to be positive. As in the original study \cite{Frank-Ginzburg}, we will consider an infinite path for the charge, i.e., its movement will occur in the interval $z'\in(-\infty,\infty)$. This means that the particle comes from the upper medium, crosses the interface and continues its path through the lower medium. As discussed in Ref. \cite{Ginzburg}, a charged particle with sufficient velocity can also emit Vavilov-\v{C}erenkov radiation in any of the two media. To focus on the physics of the transition radiation, we select the charge velocity as
 \begin{equation}\label{v choice}
v<\mathrm{min}\left\{ \frac{c}{n_1} \;,\; \frac{c}{n_2} \right\} \;,
\end{equation}
assuring that Vavilov-\v{C}erenkov radiation does not occur within any of the media, where $n_1=\sqrt{ \varepsilon_1(\omega) \mu_1(\omega) }$ denote the refractive index of medium 1 and $n_2=\sqrt{ \varepsilon_2(\omega) \mu_2(\omega) }$ that of medium 2. In a more general treatment, one would have to separate Vavilov-\v{C}erenkov and transition radiation as analyzed in the work \cite{Bass-Yakovenko} or even study the possibility of a hybrid radiation due to the interference of both kinds of radiation as investigated in Ref. \cite{Shevelev et al}.

As the current density has a single component, then the relevant components of the Green's function at both sides of the interface are only $\mathbb{G}^{iz}(\mathbf{r},\mathbf{r}';\omega)$ with $i=x,y,z$, whose explicit components are written in Appendix \ref{A}. Employing these components we perform the indicated convolution of Eq. (\ref{E GF j}) by substituting the current density (\ref{rho y j}). The involved integrals can be simplified by converting to polar coordinates, after which the angular integrals are computed analytically using integral representations of the Bessel functions. The resulting Hankel transforms of the electric field components in the reflective region, upon carrying out the integral over the  parallel wave vector magnitude  $k_\parallel=\sqrt{ k_{x}^2 + k_{y}^2 }$ and the coordinate $z'$, are
\begin{eqnarray}
E^{\,x}_1(\mathbf{r};\omega) &=& -\mathrm{i}q\omega\mu_0\mu_1(\omega)\int_{\infty}^{0}dz' \frac{e^{\mathrm{i}k_1R}}{4\pi R} \frac{ x(z-z') }{ R^2 } e^{-\mathrm{i}\omega\frac{z'}{v}} \nonumber\\
&& -\frac{\mathrm{i}q\omega\mu_0\mu_1(\omega)}{4\pi k_1^2} \frac{ x }{ R_\parallel } \frac{\partial}{\partial R_\parallel} \int_{\infty}^{0} dz' e^{-\mathrm{i}\omega\frac{z'}{v}} \mathcal{I}_1
\nonumber\\
&& +\frac{\mathrm{i}q\omega\mu_0\mu_1(\omega)}{4\pi k_1} \frac{ y }{ R_\parallel } \frac{\partial}{\partial R_\parallel} \int_{\infty}^{0} dz' e^{-\mathrm{i}\omega\frac{z'}{v}} \mathcal{I}_2
\;, \nonumber \label{Ex1 int} \\
\end{eqnarray}
\begin{eqnarray}
E^{\,y}_1(\mathbf{r};\omega) &=& -\mathrm{i}q\omega\mu_0\mu_1(\omega)\int_{\infty}^{0}dz' \frac{e^{\mathrm{i}k_1R}}{4\pi R}  \frac{ y(z-z') }{ R^2 } e^{-\mathrm{i}\omega\frac{z'}{v}} \nonumber\\
&& -\frac{\mathrm{i}q\omega\mu_0\mu_1(\omega)}{4\pi k_1^2} \frac{ y }{ R_\parallel } \frac{\partial}{\partial R_\parallel} \int_{\infty}^{0} dz' e^{-\mathrm{i}\omega\frac{z'}{v}} \mathcal{I}_1
 \nonumber\\
&& -\frac{\mathrm{i}q\omega\mu_0\mu_1(\omega)}{4\pi k_1} \frac{ x }{ R_\parallel } \frac{\partial}{\partial R_\parallel} \int_{\infty}^{0} dz' e^{-\mathrm{i}\omega\frac{z'}{v}} \mathcal{I}_2
\;, \nonumber \label{Ey1 int}\\
\end{eqnarray}

\begin{eqnarray}
E^{\,z}_1(\mathbf{r};\omega) &=& -\mathrm{i}q\omega\mu_0\mu_1(\omega) \nonumber\\
&& \times \int_{\infty}^{0}dz' \frac{e^{\mathrm{i}k_1R}}{4\pi R}\left[1 - \frac{(z-z')^2 }{ R^2 }\right] e^{-\mathrm{i}\omega\frac{z'}{v}}  \nonumber\\
&& -\frac{q\omega\mu_0\mu_1(\omega)}{4\pi k_1^2}  \int_{\infty}^{0}dz' e^{-\mathrm{i}\omega\frac{z'}{v}} \mathcal{I}_3
\;, \label{Ez1 int} 
\end{eqnarray}
where $R_\parallel=\sqrt{ x^2 + y^2 } $, $R^2= x^2 + y^2 + (z-z')^2$ and the following integrals were defined
\begin{equation}
\mathcal{I}_1  =  \int_0^\infty k_\parallel dk_\parallel R_{TM,TM}^{12}(k_\parallel) J_0(k_\parallel R_\parallel) e^{\mathrm{i}k_{z,1}(|z|+|z'|)} \;, \label{integral I1 napp} 
\end{equation}
\begin{equation}
\mathcal{I}_2  =  \int_0^\infty \frac{ k_\parallel dk_\parallel }{ k_{z,1} } R_{TE,TM}^{12}(k_\parallel) J_0(k_\parallel R_\parallel) e^{\mathrm{i}k_{z,1}(|z|+|z'|)} \;, \label{integral I2 napp}
\end{equation}
\begin{equation}
\mathcal{I}_3  =  \int_0^\infty \frac{ k_\parallel^3 dk_\parallel }{ k_{z,1} } R_{TM,TM}^{12} J_0(k_\parallel R_\parallel) e^{\mathrm{i}k_{z,1}(|z|+|z'|)} \;. 
\end{equation}
Analogously for the transmissive region, we have
\begin{eqnarray}
E^{\,x}_2(\mathbf{r};\omega) &=& -\frac{\mathrm{i}q\omega\mu_0\mu_2(\omega)}{4\pi k_2^2} \frac{ x }{ R_\parallel } \frac{\partial}{\partial R_\parallel} \int_{0}^{-\infty}dz' e^{-\mathrm{i}\omega\frac{z'}{v}} \mathcal{I}_4
\nonumber\\
&& +\frac{\mathrm{i}q\omega\mu_0\mu_2(\omega)}{4\pi k_2} \frac{ y }{ R_\parallel } \frac{\partial}{\partial R_\parallel} \int_{0}^{-\infty} dz' e^{-\mathrm{i}\omega\frac{z'}{v}} 
\mathcal{I}_5 \;, \nonumber \label{Ex2 int}\\
\end{eqnarray}
\begin{eqnarray}
E^{\,y}_2(\mathbf{r};\omega) &=& -\frac{\mathrm{i}q\omega\mu_0\mu_2(\omega)}{4\pi k_2^2} \frac{ y }{ R_\parallel } \frac{\partial}{\partial R_\parallel} \int_{0}^{-\infty}dz' e^{-\mathrm{i}\omega\frac{z'}{v}} 
\mathcal{I}_4 \nonumber\\
&& -\frac{\mathrm{i}q\omega\mu_0\mu_2(\omega)}{4\pi k_2} \frac{ x }{ R_\parallel } \frac{\partial}{\partial R_\parallel} \int_{0}^{-\infty} dz' e^{-\mathrm{i}\omega\frac{z'}{v}} 
\mathcal{I}_5 \;, \nonumber\label{Ey2 int}\\
\end{eqnarray}
\begin{eqnarray}
E^{\,z}_2(\mathbf{r};\omega) &=& - \frac{q\omega\mu_0\mu_2(\omega)}{4\pi k_2^2} \int_{0}^{-\infty} dz' e^{-\mathrm{i}\omega\frac{z'}{v}} \mathcal{I}_6
\;, \label{Ez2 int}
\end{eqnarray}
where the following integrals were also defined
\begin{eqnarray}
\mathcal{I}_4 & = & \int_0^\infty k_\parallel dk_\parallel T_{TM,TM}^{12} J_0(k_\parallel R_\parallel) e^{\mathrm{i}k_{z,2}(|z|+|z'|)}  \,, \\ 
\mathcal{I}_5 & = & \int_0^\infty \frac{ k_\parallel dk_\parallel }{ k_{z,2} } T_{TE,TM}^{12} J_0(k_\parallel R_\parallel) e^{\mathrm{i}k_{z,2}(|z|+|z'|)} \,, \\ 
\mathcal{I}_6 & = & \int_0^\infty \frac{ k_\parallel^3 dk_\parallel }{ k_{z,2} } T_{TM,TM}^{12} J_0(k_\parallel R_\parallel) e^{\mathrm{i}k_{z,2}(|z|+|z'|)}  . 
\end{eqnarray}

In the components (\ref{Ex1 int})-(\ref{Ez1 int}) and (\ref{Ex2 int})-(\ref{Ez2 int}) terms of order higher than $R^{-2}$ were neglected and the four modified Fresnel coefficients required for the electric field components are given as follows \cite{Crosse-Fuchs-Buhmann}:
\begin{eqnarray}
R_{TM,TM}^{12} &=& \frac{(\varepsilon_{2}k_{z,1}-\varepsilon_{1}k_{z,2})\Omega_\mu + \Delta^{2}k_{z,1}k_{z,2}}{(\varepsilon_{2}k_{z,1}+\varepsilon_{1}k_{z,2})\Omega_\mu + \Delta^{2}k_{z,1}k_{z,2}} \;, \label{Fresnel 1} \\
R_{TE,TM}^{12} &=& \frac{ -2\mu_2n_1k_{z,1}k_{z,2}\Delta}{(\varepsilon_{2}k_{z,1}+\varepsilon_{1}k_{z,2})\Omega_\mu+\Delta^{2}k_{z,1}k_{z,2}} \;, \label{Fresnel 2} \\
T_{TM,TM}^{12} &=& \frac{n_2}{n_1} \frac{2\varepsilon_{1}k_{z,1}\Omega_\mu }{ (\varepsilon_{2}k_{z,1}+\varepsilon_{1}k_{z,2})\Omega_\mu + \Delta^{2}k_{z,1}k_{z,2}} ,\;\; \label{Fresnel 3} \\
T_{TE,TM}^{12} &=& \frac{ -2\mu_2n_1k_{z,1}k_{z,2}\Delta}{(\varepsilon_{2}k_{z,1}+\varepsilon_{1}k_{z,2})\Omega_\mu + \Delta^{2}k_{z,1}k_{z,2}} \;, \label{Fresnel 4}
\end{eqnarray}
where $\Omega_\mu=\mu_1 \mu_2 ( k_{z,1}\mu_2 + k_{z,2}\mu_1)$ and we recall that $\Delta$ was defined in Eq. (\ref{DELTA}) and specified for TIs in Eq. (\ref{DELTA_1}). For these modified Fresnel coefficients the notation for the superscripts described for the usual Fresnel coefficients (\ref{Usual Fresnel}) is used, while the subscripts show explicitly a mixture between polarizations TE and TM \cite{Crosse-Fuchs-Buhmann}. Naturally, the coefficients (\ref{Fresnel 1})-(\ref{Fresnel 4}) reduce correctly to the standard ones when $\Delta=0$.

Our next step is to solve the integrals over $k_\parallel$. Focusing on the radiation emitted by the particle, we will implement the far-field approximation. To this aim we employ a steepest descent method  \cite{Banhos,Chew,Mandel, Chew art, Wait, OJF-LFU}. By means of this method, whose application with full details to the involved integrals is in Appendix \ref{B}, we obtain the following results after dropping terms of order higher than $r^{-2}$ for the electric field due to reflection
\begin{eqnarray}
&& \mathbf{E}_1 (\mathbf{r};\omega) = \frac{ \beta n_1 q\omega\mu_0\mu_1(\omega) e^{\mathrm{i}k_1 r}}{4\pi k_1 r}\sin\theta_1 \nonumber\\
&& \times \left\{ \mathbf{e}_{\theta1} \left[ \frac{ 1 }{ 1 + \beta n_1 \cos\theta_1 }  + \frac{ R_{TM,TM}^{12}(\theta_1,\Delta) }{ 1 - \beta n_1 \cos\theta_1 } \right] \right. \nonumber\\
&& \left. + \mathbf{e}_{\phi} \frac{ R_{TE,TM}^{12}(\theta_1,\Delta) }{ 1 - \beta n_1 \cos\theta_1 } \right\} \nonumber\\
&& +H(\theta_1-\theta_1^{disc}) \frac{ q \mu_0 \mu_1(\omega) n_2^2 v k_2 e^{\mathrm{i}\pi/4} }{ 4\pi n_1 \left( k_2 r \sin\theta_1 \right)^2 } \nonumber\\
&& \times \frac{ \left( \mathrm{i} - \frac{ n_2\cot\theta_1 }{ \sqrt{ n_2^2 - n_1^2 }  } \right)^{-3/2} }{ 1 + \beta \sqrt{ n_2^2 - n_1^2 } }e^{ \mathrm{i}k_2r\sin\theta_1 - \sqrt{k_2^2 - k_1^2 } r\cos\theta_1 } \nonumber\\
&&\times \left[ -\frac{ \tilde{R}_{TM,TM}^{12} }{ n_1 } \left( \mathbf{e}_{\rho} - \frac{ n_2 \,\mathbf{e}_{z} }{ \sqrt{ n_1^2-n_2^2 } } \right) - \frac{ \tilde{R}_{TE,TM}^{12} }{ \sqrt{ n_1^2 - n_2^2} }\;\mathbf{e}_{\phi} \right] . \nonumber \\ \label{E reflex}
\end{eqnarray}
For the transmissive one, we finally have
\begin{eqnarray}
&&\mathbf{E}_2 (\mathbf{r};\omega) = \frac{ \beta n_2 q\omega\mu_0\mu_2(\omega) e^{\mathrm{i}k_2 r}}{4\pi k_2 r} \sin\theta_2 \nonumber\\
&& \times \left\{ \mathbf{e}_{\theta2} \frac{ T_{TM,TM}^{12}(\theta_2,\Delta) }{ 1 + \beta n_2 \cos\theta_2 } +  \mathbf{e}_{\phi} \frac{ T_{TE,TM}^{12}(\theta_2,\Delta) }{ 1 + \beta n_2 \cos\theta_2 }\right\} \nonumber\\
&& + H(\theta_2-\theta_2^{disc}) \frac{ q \mu_0 \mu_2(\omega) n_1^2 v k_1 e^{\mathrm{i}\pi/4} }{ 4\pi n_2 \left( k_1 r \sin\theta_2 \right)^2 } \nonumber\\
&& \times \frac{ \left( \mathrm{i} + \frac{ n_1\cot\theta_2 }{ \sqrt{ n_1^2 - n_2^2 }  } \right)^{-3/2} }{ 1 + \beta \sqrt{ n_1^2 - n_2^2 } }e^{ \mathrm{i}k_1r\sin\theta_2 - \sqrt{k_1^2 - k_2^2 } r\cos\theta_2 } \nonumber\\
&& \times \left[ \frac{ \tilde{T}_{TM,TM}^{12} }{ n_2 } \left( \mathbf{e}_{\rho} + \frac{ n_1 \,\mathbf{e}_{z} }{ \sqrt{ n_2^2-n_1^2 } } \right) + \frac{ \tilde{T}_{TE,TM}^{12} }{ \sqrt{ n_2^2 - n_1^2 } } \;\mathbf{e}_{\phi} \right] , \nonumber\\ \label{E trans} 
\end{eqnarray}
where $\beta=v/c$, $r=\sqrt{ x^2+y^2+z^2 }$, $\mathbf{e}_{\phi}=(-\sin\phi,\cos\phi,0)$ with $\phi$ denoting the azimuthal angle, $\mathbf{e}_{\rho}=(\cos\phi,\sin\phi,0)$, $\theta_1\in[0,\pi/2)$ and $\theta_2[0,\pi/2)$ are the incidence and refraction angles respectively which define the unitary vectors $\mathbf{e}_{\theta1,2}=(\cos\theta_{1,2}\cos\phi,\cos\theta_{1,2}\sin\phi,-\sin\theta_{1,2})$ at both sides of the interface.  Both angles are measured from the normal to the interface as Snell's law dictates and $\theta_1$ coincides with the zenith angle of the observer. The integrations over $z'$ have been carried out and are translated in the \v{C}erenkov denominators $1\pm\beta n_{1,2} \cos\theta_{1,2}$ inside the keys, where for absorbing media, the condition $\mathrm{Im}(n_1),\mathrm{Im}(n_2)>0$ ensures a convergent integral over $z'$. Also in the electric fields (\ref{E reflex}) and (\ref{E trans}) the following coefficients have been introduced
\begin{eqnarray}
\tilde{R}_{TM,TM}^{12} &=& \frac{ n_2^2 \mu_1^2 - n_1^2 \mu_2^2 + \Delta^2 }{ n_2 \mu_1 \mu_2 \sqrt{ n_1^2 - n_2^2} } \;, \label{Fresnel crossed 1}\\
\tilde{R}_{TE,TM}^{12} &=& - \frac{ 2n_1 \Delta }{ n_2 \mu_1 \sqrt{ n_1^2 - n_2^2} } \;, \label{Fresnel crossed 2}\\
\tilde{T}_{TM,TM}^{12} &=& - \frac{ 2n_2 }{ \sqrt{ n_2^2 - n_1^2 } } \;, \\
\tilde{T}_{TE,TM}^{12} &=& - \frac{ 2\Delta }{ \mu_1 \sqrt{ n_2^2 - n_1^2 } } \;, \label{Fresnel crossed 4}
\end{eqnarray}
which arose as consequence of the approximation of the integrals near to the interface. Physically the coefficients (\ref{Fresnel crossed 1})-(\ref{Fresnel crossed 4}) account for the reflected and transmitted surface waves measured by an observer located close to the interface. For full details, please see Appendix \ref{B}. 

We have verified that the electric fields (\ref{E reflex}) and (\ref{E trans}) satisfy the boundary conditions (\ref{BC1}) and (\ref{BC2}). At this point is necessary to discuss the physical meaning of the electric field given by Eqs. (\ref{E reflex}) and (\ref{E trans}). First of all, we can verify that $\mathbf{e}_{r 1,2}\cdot\mathbf{E}_{1,2}=0$ as required for the spherical contribution, with $\mathbf{e}_{r 1,2}=(\sin\theta_{1,2}\cos\phi,\sin\theta_{1,2}\sin\phi,\cos\theta_{1,2})$. Secondly, we observe that the electric field in both media consists of a sum of two kind of waves: spherical and lateral. Let us describe the first of them. The spherical wave contribution decays as $r^{-1}$ as usual, nevertheless it can be split into two terms: one associated to the free space contribution plus one proportional to the modified Fresnel coefficient $R_{TM,TM}^{12}$ in the reflective region or $T_{TM,TM}^{12}$ in the transmissive one that includes a quadratic contribution of the topological parameter $\Delta$, as can be seen from Eqs. (\ref{Fresnel 1}) and (\ref{Fresnel 3}). This first contribution oscillates throughout the $\mathbf{e}_{\theta}$ direction of the corresponding medium. Then, the second term proportional to the mixed coefficient ($R_{TE,TM}^{12}$ in medium 1 or $T_{TE,TM}^{12}$ in medium 2) tells us that it has a topological origin, because it does not exist in the standard case ($\Delta=0$). Moreover, this second part of the spherical wave contribution is linear in the topological parameter $\Delta$, as Eqs. (\ref{Fresnel 2}) and (\ref{Fresnel 4}) show, and due to its vectorial dependence it only oscillates in the $\mathbf{e}_{\phi}$ direction parallel to the interface. 

Besides the spherical contribution we find the lateral wave contribution. This contribution is modulated by the corresponding Heaviside function of the medium, which discriminates if the lateral wave should be neglected or not by means of what we call the discarding angle. For the medium 1, the discarding angle $\theta_1^{disc}$ located at the upper hemisphere is given by \cite{Banhos}
\begin{equation} \label{theta UH}
\theta_1^{disc} = \frac{ \pi }{ 4 } + \frac{ n_1 }{ n_2 } + \frac{ 1 }{ 4 } \left( \frac{ n_1 }{ n_2 } \right)^2 + \mathcal{ O }\left( \frac{ n_1 }{ n_2 } \right)^3 \;.
\end{equation}
The discarding angle $\theta_2^{disc}$ for medium 2 at the lower hemisphere is given by \cite{Banhos}
\begin{equation} \label{theta LH}
\theta_2^{disc} = \frac{ n_1 }{ n_2 } \left( \sqrt{2} -1 \right)+ \mathcal{ O }\left( \frac{ n_1 }{ n_2 } \right)^3 \;,
\end{equation}
which is measured from the interface between the media. Closer inspection of Eqs. (\ref{theta UH}) and (\ref{theta LH}) shows that they do not depend on the topological parameter $\Delta$. This is due to the fact that they are obtained through the branching points defined by $k_{z,2}=\sqrt{ k_2^2-k_1^2\sin^2\theta_1 }=0$ in the upper hemisphere and by $k_{z,1}=\sqrt{ k_1^2-k_2^2\sin^2\theta_1 }=0$ in the lower one, which clearly do not depend on $\Delta$. Physically speaking such angles do not depend on the topological parameter $\Delta$ because they are characterized by the refractive indexes of both media. Thus they are defined by bulk properties, which are unaffected by the topological parameter $\Delta$. The full derivation of both discarding angles  $\theta_1^{disc}$ and $\theta_2^{disc}$ can be found in Ref. \cite{Banhos}.

Turning now to the functional dependence of the lateral waves, we observe an algebraic decay $1/\rho^2=1/(r\sin\theta_{1,2})^2$ for observation points near to the interface. The angular dependence of these lateral waves is the same of those generated by a vertical dipole \cite{Chew}, which can be expected because the medium becomes polarized due to the moving charge. Also we appreciate, in the $\mathbf{e}_{\rho}$ direction, that the phase velocity of these waves corresponds to the one of the opposite medium. In addition to the angular restriction that the Heaviside function imposes, the lateral wave contribution of the electric fields (\ref{E reflex}) decays exponentially away from the interface if $n_2>n_1$ and conversely for Eq. (\ref{E trans}). Since these waves are only observed close to the surface, they are surface waves. Furthermore, the fields (\ref{E reflex}) and (\ref{E trans}) show two terms with different directions of oscillation. The first one along the $\mathbf{e}_{\rho}$ and $\mathbf{e}_{z}$ directions is associated to the coefficients $\tilde{R}_{TM,TM}^{12}$ in medium 1 or $\tilde{T}_{TM,TM}^{12}$ in medium 2. More interesting results the second one that oscillates along the $\mathbf{e}_{\phi}$ direction. We appreciate its proportionality to the crossed coefficients $\tilde{R}_{TE,TM}^{12}$ or $\tilde{T}_{TE,TM}^{12}$, due to this dependence it becomes linear in the topological parameter $\Delta$, as Eqs. (\ref{Fresnel crossed 2}) and (\ref{Fresnel crossed 4}) show. Moreover, it has a topological origin, because it does not exist in the standard case ($\Delta=0$).

To conclude this section, we would like to give two final comments. First, we would like to remark that lateral waves have been studied intensely in the literature \cite{Banhos, Chew, Brekhovskikh, King} and their mathematical origin has been clearly understood, concretely, it comes from the fact that all the integrals over $k_\parallel$ involved in Eqs. (\ref{Ex1 int})-(\ref{Ez2 int}) only have branch-point singularities. And finally, despite the full mathematical meaning of these waves whose full details are in Appendix \ref{B}, they can be interpreted physically for an observer in medium 1 as a part of a spherical wave in medium 2 that is propagating along the interface, and is being refracted back as evanescent waves \cite{Chew}.

%
\section{Angular Distribution of Radiation}\label{ANGULAR}

In this section, we will determine the angular distribution of radiation in the upper hemisphere. Henceforth, we will focus only on the spherical wave contribution meaning, according to our results, that our materials choice should provide a suitable discarding angle $\theta_1^{disc}$ (\ref{theta UH}) in order to assure the absence of the surface waves.

For the purpose of the current section, we need to express the electric field of the lower hemisphere in terms of the variables that describe the upper one \cite{Frank-Ginzburg, Frank Lecture}. This can be done by employing Snell's law, valid for three-dimensional TIs and magnetoelectric media \cite{Chang-Yang,Obukhov-Hehl},  which allows us to write the full electric field in terms of $\theta_1$ in the following compact way:
\begin{eqnarray}
&& \mathbf{E}_1 (\mathbf{r};\omega) = \frac{ \beta n_1 q\omega\mu_0 e^{\mathrm{i}k_1 r}}{4\pi k_1 r}\sin\theta_1 \nonumber\\
&& \times \left\{ \mathbf{e}_{\theta1} \left[ \frac{ \mu_1(\omega) }{ 1 + \beta n_1 \cos\theta_1 } + \frac{ \mu_1(\omega) R_{TM,TM}^{12}(\theta_1,\Delta) }{ 1 - \beta n_1 \cos\theta_1 } \right. \right. \nonumber\\
&& \left. -  \frac{ n_1 \mu_2(\omega) T_{TM,TM}^{12}(\theta_1,\Delta) }{ n_2 \left (1 + \beta n_2\cos\theta_2 \right) }  \right]  \nonumber\\
&& +  \mathbf{e}_{\phi} \left[ \frac{ \mu_1(\omega) R_{TE,TM}^{12}(\theta_1,\Delta) }{ 1 - \beta n_1 \cos\theta_1 } \right. \nonumber\\
&& \left. \left. - \frac{ n_1 \mu_2(\omega) T_{TE,TM}^{12}(\theta_2,\Delta) }{ n_2 \left( 1 + \beta n_2\cos\theta_2 \right) } \right] \right\}  \;, \label{E reflex whole}
\end{eqnarray}
where $n_2\cos\theta_2=\sqrt{ n_2^2 - n_1^2 \sin^2\theta_1}$. In this final expression for the electric field, the four modified Fresnel coefficients given in Eqs. (\ref{Fresnel 1})-(\ref{Fresnel 4})  have been converted to angular functions through the substitutions $k_{z,1}= k_1 \cos\theta_1$ $k_{z,2}=\sqrt{ k_2^2 - k_1^2 \sin^2\theta_1 }$, as a consequence of the steepest descent method employed to approximate the involved integrals.  

Using Maxwell equations to leading order in $1/r$, we verify the expressions 
\begin{equation} \label{Radiation Fields}
\mathbf{e}_{r1}\cdot\mathbf{E}_1=0\;, \quad \mathbf{e}_{r1}\cdot\mathbf{B}_1=0\;, \quad \mathbf{B} = \mathbf{e}_{r1} \times ( n_1 \mathbf{E}_1 ) \;,
\end{equation}
which are the distinctive feature of the radiation fields \cite{Jackson}, with $n_1=\sqrt{\varepsilon_1 \mu_1}$. We observe that the three vectors $\mathbf{E}_1$, $\mathbf{B}_1$, and $\mathbf{e}_{r1}$ define a right-handed triad resulting in the Poynting vector for the material medium 1. Then, the angular spectral density of the radiation over the solid angle $\Omega$ can be defined through the Poynting vector, but recalling expressions (\ref{Radiation Fields}) it can be defined in the following fashion \cite{Schwinger}:
\begin{equation} \label{Angular distribution}
\frac{ d^2 \mathcal{E}_1 }{ d\omega d\Omega } = \sqrt{ \frac{ \varepsilon_0\varepsilon_1 }{ \mu_0 \mu_1} } \frac{ r^2 }{ \pi } \mathbf{E}_1^*(\mathbf{r};\omega) \cdot \mathbf{E}_1(\mathbf{r};\omega) \;, \\
\end{equation}
which for our configuration leads to
\begin{eqnarray}
&& \frac{ d^2 \mathcal{E}_1 }{ d\omega d\Omega } = \frac{ q^2 }{ 4\pi\varepsilon_0 } \sqrt{ \frac{ \varepsilon_1 }{ \mu_1 }} \frac{ v^2 }{ 4\pi^2 c^3 } \sin^2\theta_1 \left\{ \left[ \frac{ \mu_1 }{ 1 + \beta n_1 \cos\theta_1 } \right. \right. \nonumber\\
&& \left. + \frac{ \mu_1 R_{TM,TM}^{12}(\theta_1,\Delta) }{ 1 - \beta n_1 \cos\theta_1 } -  \frac{ n_1 \mu_2 T_{TM,TM}^{12}(\theta_1,\Delta) }{ n_2 \left (1 + \beta  n_2\cos\theta_2 \right) } \right]^2  \nonumber\\
&& \left. + \left[ \frac{ \mu_1  R_{TE,TM}^{12}(\theta_1,\Delta) }{ 1 - \beta n_1 \cos\theta_1 }  - \frac{ n_1 \mu_2 T_{TE,TM}^{12}(\theta_1,\Delta) }{ n_2 \left (1 + \beta n_2\cos\theta_2 \right) }\right]^2  \right\} \;, \nonumber\\ \label{Angular distribution TR} 
\end{eqnarray}
where we assumed frequency-independent values for $\varepsilon_1, \mu_1, \varepsilon_2$ and $\mu_2$ in order to apply the formula (\ref{Angular distribution}) as well in what follows of the current work. The subindex of $\mathcal{E}_1$ denotes that the observer is located at the upper hemisphere and the particle comes from medium 1 towards medium 2.

Some comments regarding this angular distribution are now in order. The expression (\ref{Angular distribution TR}) has no linear contribution in the topological parameter $\Delta$ and is an even function of the angle $\phi$, the last fact being a consequence of the azimuthal symmetry of the problem (recall Fig. \ref{REGIONS}). The first term consists in a sum of three terms that represent: (i) the field of the charged particle $q$ moving in medium 1 until reaching the origin at the interface of the medium 2. (ii) The field of the charged particle $q$ moving from the origin into the medium 2. (iii) The field of the electrical image of the particle moving from the interior of medium 2 towards the origin at the interface \cite{Frank-Ginzburg, Frank Lecture}. This situation is the same as in the standard electrodynamics situation but the Fresnel coefficients have additional contributions due to the topological parameter, see Eqs. (\ref{Fresnel 1}) and ({\ref{Fresnel 3}). These contributions modify the image electric charge and have the consequence that the well-known relationship $\varepsilon_1n_2(1+r^{12}_{TM}) = \varepsilon_2 n_1 t^{12}_{TM}$ \cite{Buhmann 1} for the standard Fresnel coefficients does not hold for these materials, see discussion of Sec. \ref{EFT}. Furthermore, Eq. (\ref{Angular distribution TR}) has a second term whose origin is completely topological, as can be appreciated from the dependence of the mixed Fresnel terms $R_{TE,TM}^{12}$ (\ref{Fresnel 2}) and $T_{TE,TM}^{12}$ (\ref{Fresnel 4}). Interestingly, this second term can be interpreted qualitatively by applying the ideas of fields generated by dynamical images of the work \cite{OJF-LFU-ORT}, which extended the original ones developed in the articles \cite{Qi Science,Wilczek}. In our case we will have a superposition of the electric field  of two image magnetic monopoles with same strengths but opposite signs: (i) One image magnetic monopole $g_1=qR_{TE,TM}^{12}\sim -q\Delta\sim -q\alpha$ starting at the same moment as the electric charge $q$ from the point origin going into the medium 2. (ii) And another one with strength $g_2=qT_{TE,TM}^{12}\sim q\Delta\sim q\alpha$ moving from the interior of medium 2 towards the origin at the interface and stopping at that point. 

Setting $\mu_1=\mu_2=1$ and $\Delta=0$, i.e. turning off the topological parameter or the magnetoelectric coupling constant and for non-magnetic media, Eq. (\ref{Angular distribution TR}) reproduces the original result by Frank and Ginzburg  \cite{Frank-Ginzburg, Frank Lecture} for two different dielectric media: 
\begin{eqnarray}
&& \frac{ d^2 \mathcal{E}_1 }{ d\omega d\Omega } = \frac{ q^2 \sqrt{ \varepsilon_1 } }{ 4\pi\varepsilon_0 } \frac{ v^2 }{ 4\pi^2 c^3 } \sin^2\theta_1 \left[ \frac{ 1 }{ 1 + \beta n_1 \cos\theta_1 } \right. \nonumber\\
&& \left. + \frac{ r_{TM}^{12}(\theta_1) }{ 1 - \beta n_1 \cos\theta_1 } -  \frac{ n_1 t_{TM}^{12}(\theta_1) }{ n_2 \left (1 + \beta \sqrt{ n_2^2 - n_1^2\sin^2\theta_1 } \right) } \right]^2 \;, \nonumber\\
\end{eqnarray}
where $r_{TM}^{12}(\theta_1)=R_{TM,TM}^{12}(\theta_1,0)$ and $t_{TM}^{12}(\theta_1)=T_{TM,TM}^{12}(\theta_1,0)$ denote the standard Fresnel coefficients for the TM polarization.

At an interface made up of an ideal conductor and an upper nonmagnetic medium with constant permittivity $\varepsilon_1$, the expression (\ref{Angular distribution TR}) reproduces the Frank and Ginzburg formula \cite{Frank-Ginzburg, Frank Lecture}
\begin{equation}\label{Angular distribution TR Ideal}
\frac{ d^2 \mathcal{E}_1 }{ d\omega d\Omega } = \frac{ q^2 \sqrt{ \varepsilon_1 } }{ 4\pi\varepsilon_0 } \frac{ v^2 }{ 4\pi^2 c^3 } \frac{ \sin^2\theta_1 }{ (1 - \beta^2 n_1^2 \cos^2\theta_1 )^2 } \;,
\end{equation}
which constitutes another consistency check for our results.

Equation (\ref{Angular distribution TR}) can be easily approximated for non-relativistic particles ($\beta\ll1$)
\begin{eqnarray}
&& \frac{ d^2 \mathcal{E}_1 }{ d\omega d\Omega }  =  \frac{ q^2 }{ 4\pi\varepsilon_0 } \sqrt{ \frac{ \varepsilon_1 }{ \mu_1 }} \frac{ v^2 }{ 4\pi^2 c^3 } \sin^2\theta_1 \left\{ \left[ \; \mu_1 \right. \right. \nonumber\\
&&   + \left. \left.  \mu_1 R_{TM,TM}^{12}(\theta_1,\Delta)  -  \frac{ n_1 \mu_2 }{ n_2 } T_{TM,TM}^{12}(\theta_1,\Delta) \right]^2 \right. \nonumber\\
&& + \left.  \left[  \mu_1  R_{TE,TM}^{12}(\theta_1,\Delta)   - \frac{ n_1 \mu_2 }{ n_2 } T_{TE,TM}^{12}(\theta_1,\Delta) \right]^2  \right\} \;. \nonumber\\
\end{eqnarray}

Before presenting plots that illustrate the behavior of the angular distribution of radiation (\ref{Angular distribution TR}), we remark the presence of the denominators of the type  $1\pm\beta n_{1}\cos\theta_1$
and $1 + \beta \sqrt{ n_2^2 - n_1^2 \sin^2\theta_1}$ that are characteristic of the radiation of a relativistic particle. Of course, from the Vavilov-\v{C}erenkov radiation we recognize that these denominators can be zero to a certain angle corresponding to the well-known \v{C}erenkov cone ray in the corresponding medium. This is described when the intensity becomes infinite. However, given our selection of the particle velocity discussed in Sec. \ref{CHARGE} the Vavilov-\v{C}erenkov conditions are not fulfilled and we will obtain a pure transition radiation phenomenon.
\begin{figure}[tbp]
\includegraphics[width=0.45\textwidth]{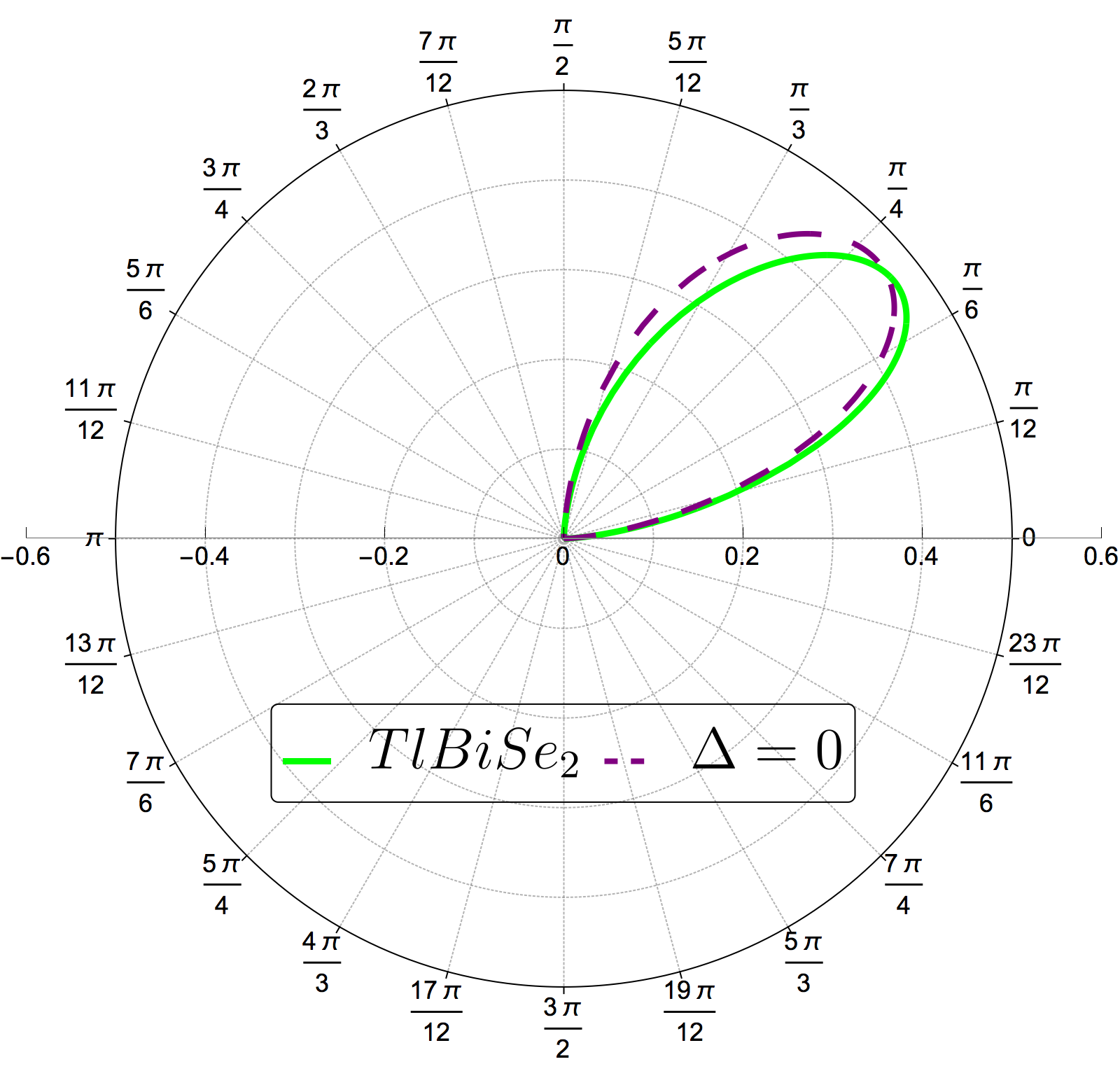}
\caption{ Comparison of the angular distribution of standard transition radiation and the pure topological term of Eq. (\ref{Angular distribution TR}) when the radiation is generated by a particle with $\lambda=3000\mathring{A}$ and two different velocities. The standard ($\Delta=0$) transition radiation for an interface constituted by vacuum and a standard dielectric medium with $\varepsilon_2=4$ and $\mu_2=1$ corresponds to the dashed purple line. The green solid line represents the pure topological term of Eq. (\ref{Angular distribution TR}) but for an interface constituted by vacuum and the topological insulator TlBiSe$_2$ with $\varepsilon_2=4$, $\mu_2=1$ and $\Delta=11\alpha$. Polar plot for $v=0.75c$, here the topological contribution was multiplied by a factor of 785.4319. The radial axis indicates the dimensionless angular distribution $\left(  q^2 v^2 / 4\pi\varepsilon_0 4\pi^2 c^3  \right)^{-1}\ d^2 \mathcal{E}_1/ d\omega d\Omega$ in the respective direction. The charge moves from right to left. }
\label{PLOTS A1}
\end{figure}
\begin{figure}[tbp]
\includegraphics[width=0.45\textwidth]{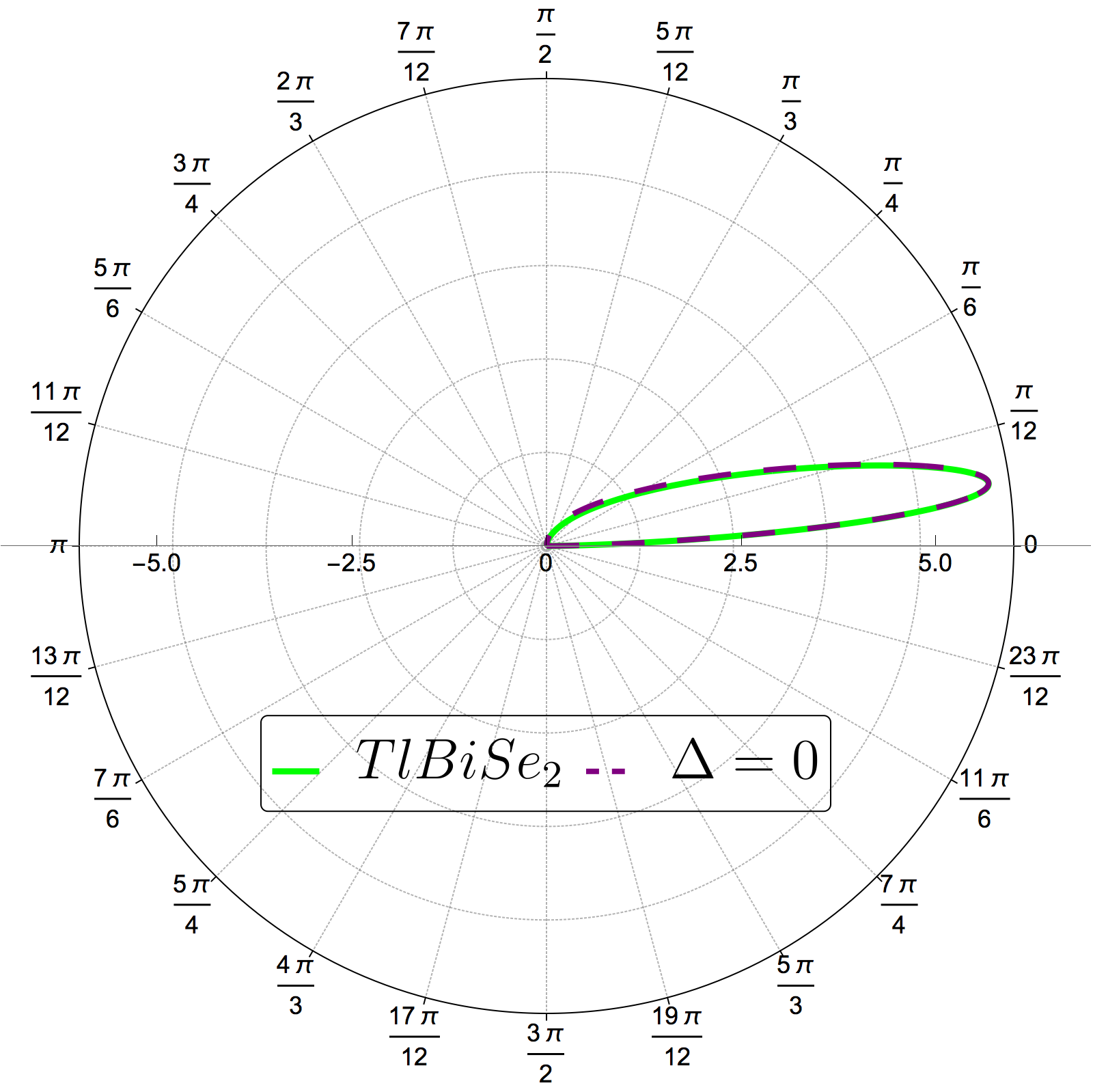}
\caption{ Comparison of the angular distribution of standard transition radiation and the pure topological term of Eq. (\ref{Angular distribution TR}) when the radiation is generated by a particle with $\lambda=3000\mathring{A}$ and two different velocities. The standard ($\Delta=0$) transition radiation for an interface constituted by vacuum and a standard dielectric medium with $\varepsilon_2=4$ and $\mu_2=1$ corresponds to the dashed purple line. The green solid line represents the pure topological term of Eq. (\ref{Angular distribution TR}) but for an interface constituted by vacuum and the topological insulator TlBiSe$_2$ with $\varepsilon_2=4$, $\mu_2=1$ and $\Delta=11\alpha$. Polar plot for $v=0.99c$, here the topological contribution was multiplied by a factor of 361.3548. The radial axis indicates the dimensionless angular distribution $\left(  q^2 v^2 / 4\pi\varepsilon_0 4\pi^2 c^3  \right)^{-1}\ d^2 \mathcal{E}_1/ d\omega d\Omega$ in the respective direction. The charge moves from right to left.  }
\label{PLOTS A2}
\end{figure}

Recalling the expressions (\ref{Fresnel 1})-(\ref{Fresnel 4}) for the modified Fresnel coefficients, the transition radiation of Eq. (\ref{Angular distribution TR}) receives additional contributions of order $\Delta^2$ and $\Delta^4$. Nevertheless, the second term with a purely topological origin is of order $\Delta^2$ and it does not have the term without any modified Fresnel coefficients, implying that it has a slightly different angular behavior with respect to the whole transition radiation phenomenon. For this reason, we present first this term in Figs. \ref{PLOTS A1} and \ref{PLOTS A2}, which contrast the behavior between such term for the topological insulator TlBiSe$_2$ and the standard ($\Delta=0$) transition radiation. Both angular distributions of radiation observed in the upper hemisphere are generated by a particle with $\lambda=3000\mathring{A}$ and two different velocities, $v=0.75c$ for Fig. \ref{PLOTS A1} and $v=0.99c$ for Fig. \ref{PLOTS A2}. Standard transition radiation for an interface constituted by vacuum and a standard dielectric medium with $\varepsilon_2=4$ and $\mu_2=1$ corresponds to the dashed purple line in Figs. \ref{PLOTS A1} and \ref{PLOTS A2}. On the other hand, the solid green lines represent the pure topological contribution of Eq. (\ref{Angular distribution TR}) discussed above but for an interface constituted by vacuum and the topological insulator TlBiSe$_2$ with $\varepsilon_2=4$, $\mu_2=1$, and $\Delta=11\alpha$ \cite{TlBiSe2}. Figures \ref{PLOTS A1} and \ref{PLOTS A2} represent the polar plots of the radiation in which the interface lies on the line ($\pi/2-3\pi/2$) and the charge moves from right to left along the line ($0-\pi$). The vacuum region is in the upper hemisphere $\theta_1\in(0,\pi/2)$ and the standard dielectric medium or the topological insulator TlBiSe$_2$ in the lower hemisphere $\theta_1\in(\pi/2,\pi)$. Though we find the characteristic lobes of transition radiation for both polar plots, we observe clearly how the angles of maximal emission of these lobes are shifted for the velocity $v=0.75c$ in Fig. \ref{PLOTS A1}. In spite of this, the shift in such angles is imperceptible when the velocity is $v=0.99c$ in Fig. \ref{PLOTS A2}, i.e. in the ultra-relativistic regime the angles of maximal emission of the lobes coincide providing the possibility of an enhancement of the whole phenomenon. To perform this comparison, we multiplied by a factor of 785.4319 the pure topological term in Fig. \ref{PLOTS A1} and by 361.3548 in Fig. \ref{PLOTS A2}. 
\begin{figure}[tbp]
\includegraphics[width=0.425\textwidth]{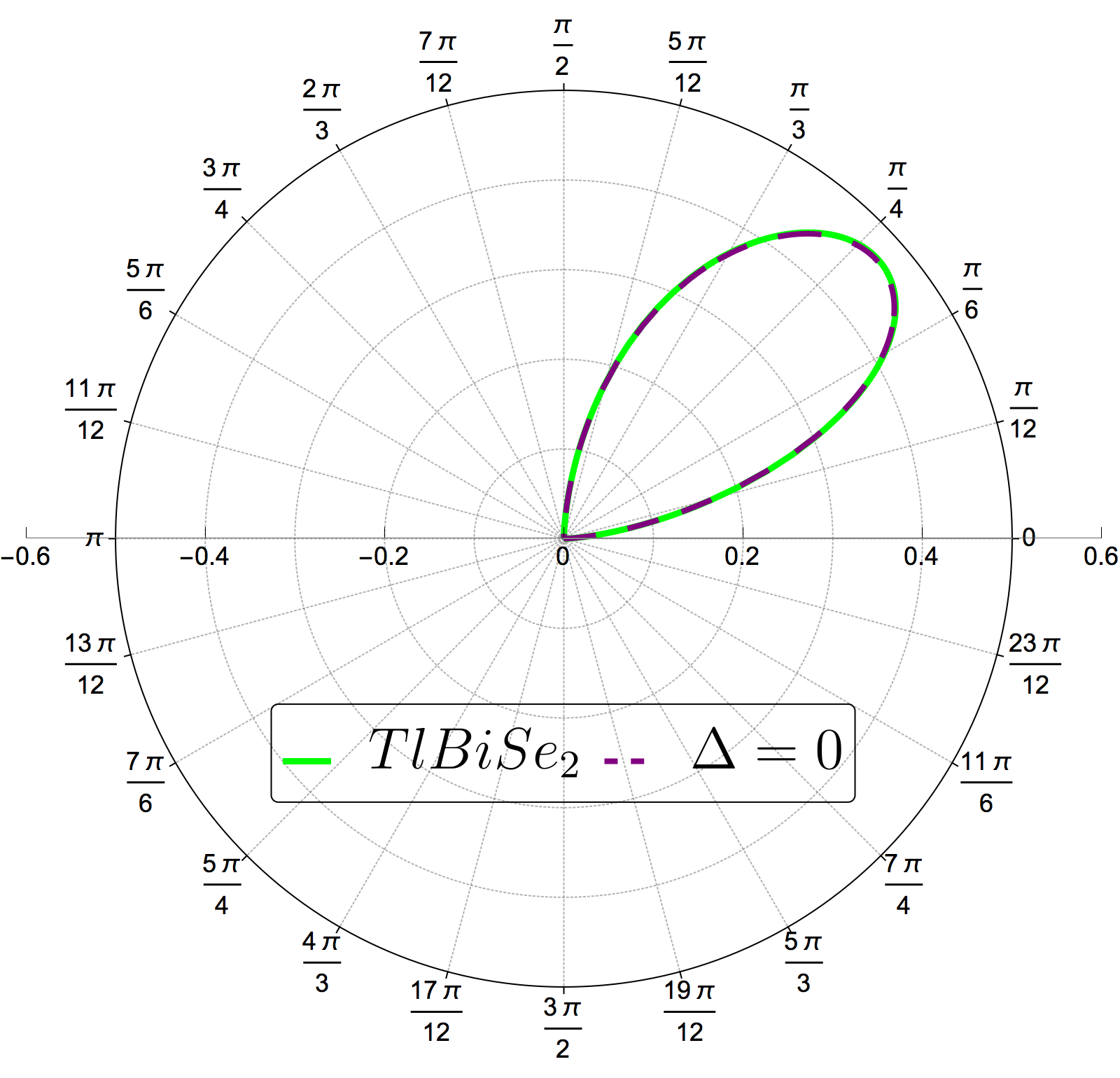}
\caption{ Angular distribution of transition radiation generated by a particle with $v=0.75c$ and $\lambda=3000\mathring{A}$. The case of standard ($\Delta=0$) transition radiation for an interface constituted by vacuum and a standard dielectric medium with $\varepsilon_2=4$ and $\mu_2=1$ corresponds to the dashed purple line. The green solid line represents the same quantity but for an interface constituted by vacuum and the topological insulator TlBiSe$_2$ with $\varepsilon_2=4$, $\mu_2=1$, and $\Delta=11\alpha$. Polar plot for transition radiation obtained by Eq. (\ref{Angular distribution TR}). The radial axis indicates the dimensionless angular distribution $\left(  q^2 v^2 / 4\pi\varepsilon_0 4\pi^2 c^3  \right)^{-1}\ d^2 \mathcal{E}_1/ d\omega d\Omega$ in the respective direction. The charge moves from right to left. }
\label{PLOTS B1}
\end{figure}
\begin{figure}[tbp]
\includegraphics[width=0.45\textwidth]{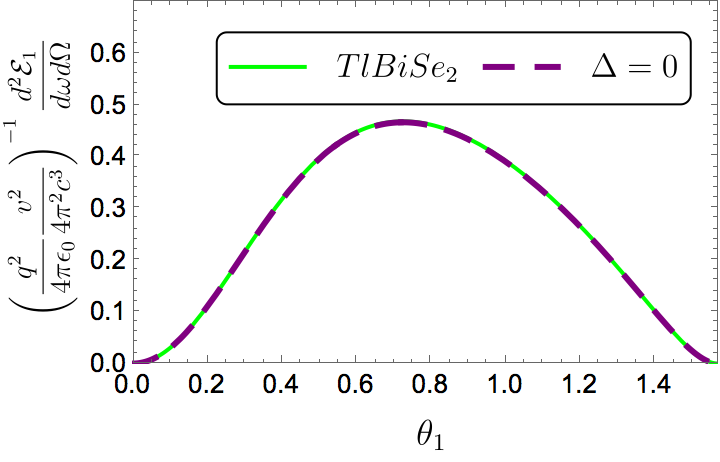}
\caption{ Angular distribution of transition radiation generated by a particle with $v=0.75c$ and $\lambda=3000\mathring{A}$. The case of standard ($\Delta=0$) transition radiation for an interface constituted by vacuum and a standard dielectric medium with $\varepsilon_2=4$ and $\mu_2=1$ corresponds to the dashed purple line. The green solid line represents the same quantity but for an interface constituted by vacuum and the topological insulator TlBiSe$_2$ with $\varepsilon_2=4$, $\mu_2=1$, and $\Delta=11\alpha$.  Standard plot for transition radiation obtained by Eq. (\ref{Angular distribution TR}). The scale is in arbitrary dimensions. The charge moves from medium 1 to medium 2.  }
\label{PLOTS B2}
\end{figure}

For sake of comparison between the standard transition radiation and all the modifications arisen from the  strong three-dimensional TI, we have plotted both situations described through Eq. (\ref{Angular distribution TR}) in Figs. \ref{PLOTS B1} and \ref{PLOTS B2}. These figures show the angular distribution of radiation for transition radiation observed in the upper hemisphere generated by a particle with $v=0.75c$ and $\lambda=3000\mathring{A}$ for the same media of Figs. \ref{PLOTS A1} and \ref{PLOTS A2}. Again the standard ($\Delta=0$) transition radiation for a standard dielectric medium with $\varepsilon_2=4$ and $\mu_2=1$ corresponds to the dashed purple line in Figs. \ref{PLOTS B1} and \ref{PLOTS B2}, meanwhile the solid green lines represent the same quantity but for the topological insulator TlBiSe$_2$ with $\varepsilon_2=4$, $\mu_2=1$, and $\Delta=11\alpha$ \cite{TlBiSe2}. Figure \ref{PLOTS B1} represents a polar plot of the radiation as well as Figs. \ref{PLOTS A1} and \ref{PLOTS A2}, so the same specifications of that figures do apply. Here we find the characteristic lobes of transition radiation and observe that the modifications due to the topological nature of the TlBiSe$_2$ are almost imperceptible. Numerically we found that these differences are of the order $10^{-3}$ for these parameters, for which $\Delta=11\alpha$ is the highest value available for TlBiSe$_2$. Figure \ref{PLOTS B2} is a standard plot of the angular distribution of radiation (\ref{Angular distribution TR}), where we could observe that its maxima coincide with the lobes' maxima of Fig. \ref{PLOTS B1}. Again the modifications due to the topological nature of the TlBiSe$_2$ are almost imperceptible supporting the behavior of Fig. \ref{PLOTS B1}. The surface waves are discarded because for such parameters Eq. (\ref{theta UH}) gives a discarding angle $\theta_1^{disc}\simeq1.3479\simeq77.2289^{\circ}$, which is far away from the lobes and the main contribution of the angular distribution.
\begin{figure}[tbp]
\includegraphics[width=0.425\textwidth]{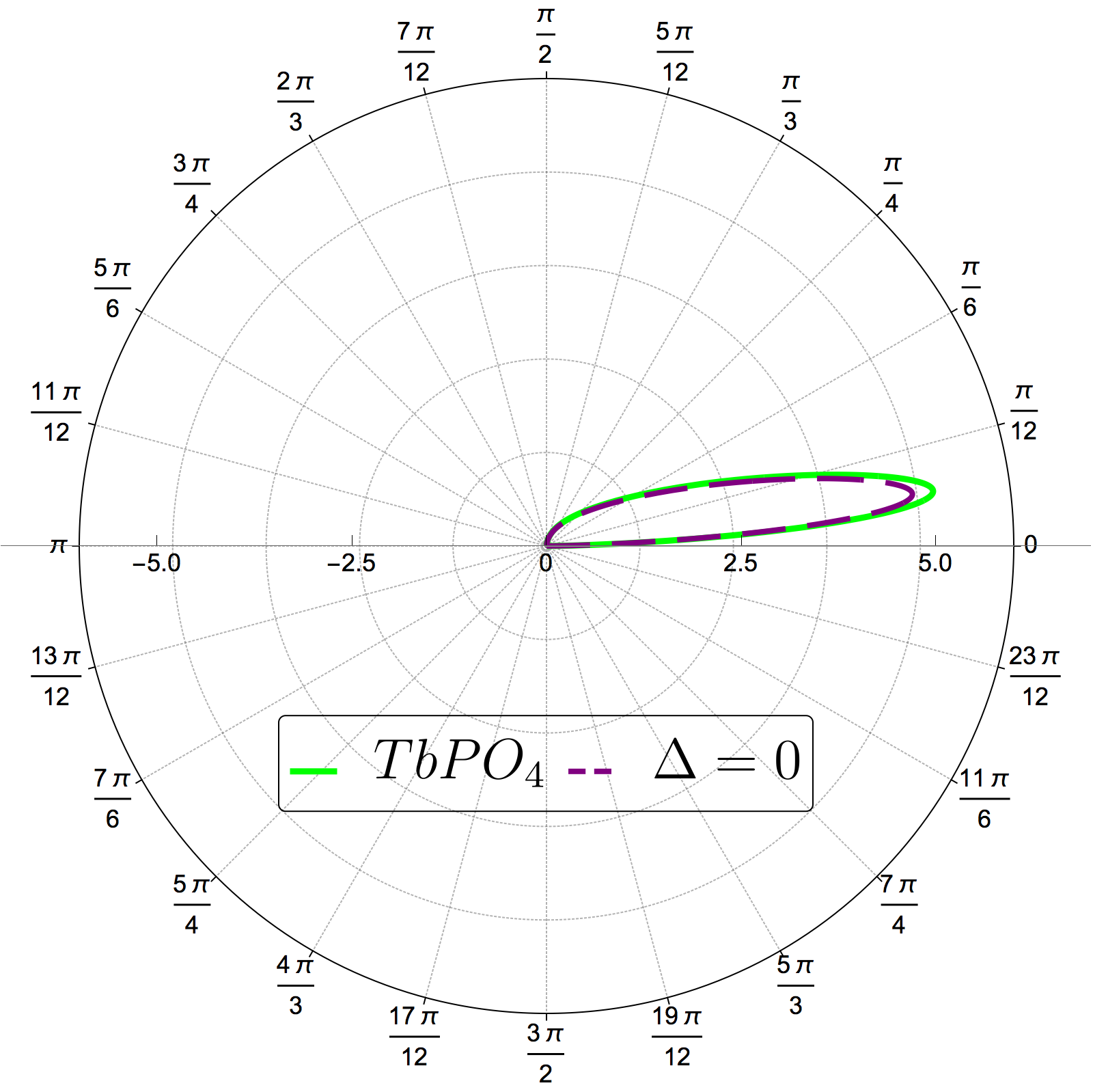}
\caption{ Angular distribution of radiation for transition radiation generated by a particle with $v=0.99c$ and $\lambda=3000\mathring{A}$. The case of standard ($\Delta=0$) transition radiation for an interface constituted by vacuum and a standard dielectric medium with $\varepsilon_2=3.4969$ and $\mu_2=1$ corresponds to the dashed purple line. The green solid line represents the same quantity but for an interface constituted by vacuum and the magnetoelectric TbPO$_4$ with $\varepsilon_2=3.4969$, $\mu_2=1$, and $\Delta=0.22$. Polar plot for transition radiation obtained by Eq. (\ref{Angular distribution TR}). The radial axis indicates the dimensionless angular distribution $\left(  q^2 v^2 / 4\pi\varepsilon_0 4\pi^2 c^3  \right)^{-1}\ d^2 \mathcal{E}_1/ d\omega d\Omega$ in the respective direction. The charge moves from right to left. }
\label{PLOTS C1}
\end{figure}
\begin{figure}[tbp]
\includegraphics[width=0.45\textwidth]{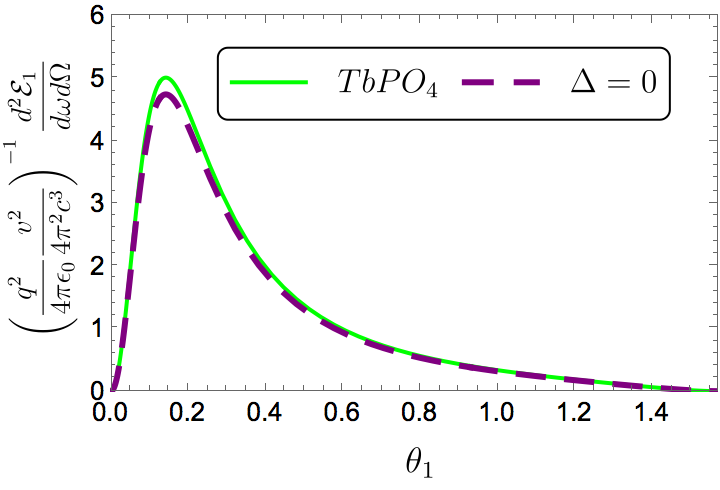}
\caption{ Angular distribution of radiation for transition radiation generated by a particle with $v=0.99c$ and $\lambda=3000\mathring{A}$. The case of standard ($\Delta=0$) transition radiation for an interface constituted by vacuum and a standard dielectric medium with $\varepsilon_2=3.4969$ and $\mu_2=1$ corresponds to the dashed purple line. The green solid line represents the same quantity but for an interface constituted by vacuum and the magnetoelectric TbPO$_4$ with $\varepsilon_2=3.4969$, $\mu_2=1$, and $\Delta=0.22$. Standard plot for transition radiation obtained by Eq. (\ref{Angular distribution TR}). The scale is in arbitrary dimensions. The charge moves from medium 1 to medium 2. }
\label{PLOTS C2}
\end{figure}

However, if we move to the ultrarelativistic regime and if we increase the topological parameter the modifications to the transition radiation become evident. Figures \ref{PLOTS C1} and \ref{PLOTS C2} illustrate this by comparing the standard transition radiation and its modifications described through Eq. (\ref{Angular distribution TR}). These figures show again the angular distribution for transition radiation observed in the upper hemisphere generated by a particle with $v=0.99c$ and $\lambda=3000\mathring{A}$. The standard ($\Delta=0$) transition radiation for an interface constituted by vacuum and a standard dielectric medium with $\varepsilon_2=3.4969$ and $\mu_2=1$ corresponds to the dashed purple line in Figs. \ref{PLOTS C1} and \ref{PLOTS C2}. On the other hand, the solid green lines represent the same quantity but for an interface constituted by vacuum and the magnetoelectric TbPO$_4$ with $\varepsilon_2=3.4969$, $\mu_2=1$ and $\Delta=0.22$ \cite{TbPO4,Rivera}. Figure \ref{PLOTS C1} represents a polar plot of the radiation and the same specifications from Figs. \ref{PLOTS A1}-\ref{PLOTS B1} apply here. Nevertheless, here we have in the lower hemisphere $\theta_1\in(\pi/2,\pi)$ the standard dielectric medium or the magnetoelectric TbPO$_4$. Again the  characteristic lobes of transition radiation emerge but now they are sharped and compressed against the polar axis. This behavior respects the typical behavior of a ultrarelativistic particle where the maximum angle of the distribution is very small being of the order of the ratio of the rest energy of the particle to its total energy \cite{Jackson}. This maximum allows to appreciate clearly the difference between the modifications due to the magnetoelectric TbPO$_4$, whose corresponding maximum is greater than the one related to the standard transition radiation. Figure \ref{PLOTS C2} is a standard plot of the angular distribution of radiation (\ref{Angular distribution TR}), where we could observe that its maxima coincide with the lobes maxima of Fig. \ref{PLOTS C1}. Also the modifications due to the magnetoelectric TbPO$_4$ are visible supporting the behavior seen in Fig. \ref{PLOTS C1}. Recalling that the particle moves from right to left along the line ($0-\pi$), this means that strong three-dimensional TIs or magnetoelectric media enhance the backward emitted radiation. The surface waves are again safely discarded because for such parameters Eq. (\ref{theta UH}) gives a discarding angle $\theta_1^{disc}\simeq1.39165\simeq79.7356^{\circ}$, which is far away of the lobes and the main behavior of the angular distribution. 
\begin{figure}[tbp]
\includegraphics[width=0.425\textwidth]{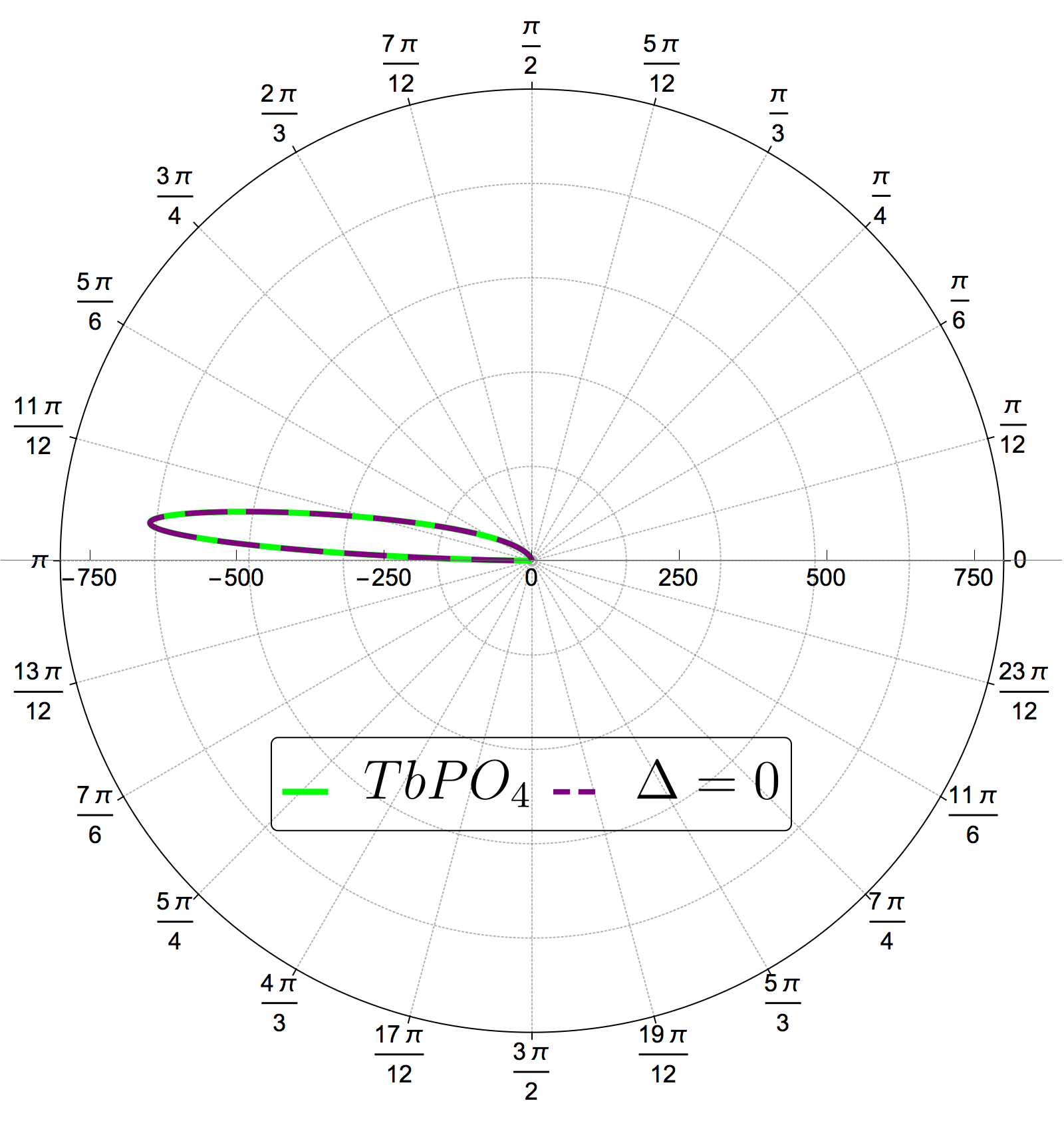}
\caption{ Angular distribution of radiation for transition radiation generated by a particle with $v=0.534c$ and $\lambda=3000\mathring{A}$. The case of standard ($\Delta=0$) transition radiation for an interface constituted by vacuum and a standard dielectric medium with $\varepsilon_2=3.4969$ and $\mu_2=1$ corresponds to the dashed purple line. The green solid line represents the same quantity but for an interface constituted by vacuum and the magnetoelectric TbPO$_4$ with $\varepsilon_2=3.4969$, $\mu_2=1$, and $\Delta=0.22$. Polar plot for transition radiation obtained by Eq. (\ref{Angular distribution TR}) when the sign of $\beta$ and the indexes 1 and 2 are reversed. The radial axis indicates the dimensionless angular distribution $\left(  q^2 v^2 / 4\pi\varepsilon_0 4\pi^2 c^3  \right)^{-1}\ d^2 \mathcal{E}_2/ d\omega d\Omega$ in the respective direction. The charge moves from left to right. }
\label{PLOTS D1}
\end{figure}
\begin{figure}[tbp]
\includegraphics[width=0.45\textwidth]{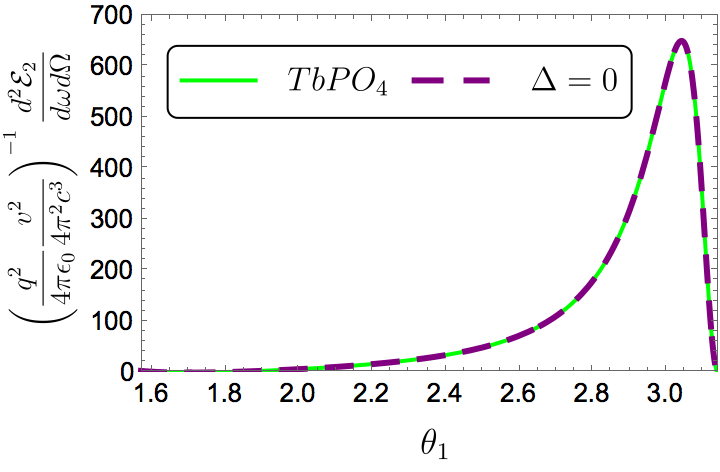}
\caption{ Angular distribution of radiation for transition radiation generated by a particle with $v=0.534c$ and $\lambda=3000\mathring{A}$. The case of standard ($\Delta=0$) transition radiation for an interface constituted by vacuum and a standard dielectric medium with $\varepsilon_2=3.4969$ and $\mu_2=1$ corresponds to the dashed purple line. The green solid line represents the same quantity but for an interface constituted by vacuum and the magnetoelectric TbPO$_4$ with $\varepsilon_2=3.4969$, $\mu_2=1$ and $\Delta=0.22$. Standard plot for transition radiation obtained by Eq. (\ref{Angular distribution TR}) when the sign of $\beta$ and the indexes 1 and 2 are reversed. The scale is in arbitrary dimensions. The charge moves from medium 2 to medium 1.  }
\label{PLOTS D2}
\end{figure}

Naturally, one can also analyze the current phenomenon for the opposite direction of the velocity (Recall Fig. \ref{REGIONS}), which means that the particle moves from medium 2 into medium 1. For such case, the sign in front of $\beta$ in all terms in the denominator of Eq. (\ref{Angular distribution TR}) should be reversed as well as the indexes 1 into 2 and vice versa. Thus the radiation patterns will depend on the sign particle velocity with respect to the medium, which will respect the directivity of the phenomenon present in standard electrodynamics. Figures \ref{PLOTS D1} and \ref{PLOTS D2} illustrate this situation for the same media used in Figs. \ref{PLOTS C1} and \ref{PLOTS C2}, in this way the same specifications for the dashed purple lines and the solid green lines apply to Figs. \ref{PLOTS D1} and \ref{PLOTS D2} too. In these last figures, we show the angular distribution of the radiation in the lower hemisphere for a charged particle moving from left to right along the line $(\pi-0)$ of Fig. \ref{PLOTS D1}. Here the particle travels with $\lambda=3000\mathring{A}$ and $v=0.534c$, whose velocity is close and below the \v{C}erenkov threshold of the media located at $v=c/1.87\simeq0.534759c$ in agreement with our velocity choice of Eq. (\ref{v choice}). Comparing Figs. \ref{PLOTS C1} and \ref{PLOTS C2} with these last ones, we observe that even for a $\Delta=0.22$ in the ultra-relativistic regime the transition radiation of standard electrodynamics is dominant and the additional contributions that depend on $\Delta$ are not significant. This will be justified quantitatively by means of the corresponding frequency distribution in the next section. Again the surface waves are neglected because for these parameters Eq. (\ref{theta LH}) gives a discarding angle $\theta_2^{disc}\simeq0.221505\simeq12.6913^{\circ}$, which means an angle of $\theta_2^{disc}\simeq3.3631\simeq102.6913^{\circ}$ for our spherical coordinate system.  

As discussed in the Introduction, TIs and magnetoelectrics can be regarded as Tellegen media and they share similar coupling of the electric and magnetic field with Pasteur media, which exhibit chirality. Thus the corresponding transition radiation will share some similarities but also exhibit important differences, which we will briefly address. Transition radiation in chiral matter has been studied in Ref. \cite{Galyamin} for an interface between vacuum and a chiral isotropic medium with the charge starting from the vacuum and crossing through that medium. Based on this work, we will compare their results with our results for the angular distribution of radiation when dissipation is negligible for both kinds of materials.

Beginning with the similarities, both electric fields propagate in the same directions $\mathbf{e}_{\theta1}$ and $\mathbf{e}_\phi$ as Eqs. (57) of Ref. \cite{Galyamin} and our Eq. (\ref{E reflex whole}) show. Particularly, for both cases the $\phi$ component is linearly proportional to the chirality or the topological parameter $\Delta$ standing for the nonreciprocity. Indeed, the nonreciprocity contribution is codified in the modified Fresnel coefficients  $R_{TE,TM}^{12}(\theta_1,\Delta)$ and the $T_{TE,TM}^{12}(\theta_1,\Delta)$ in Eq. (\ref{E reflex whole}). Regarding the radiation patterns, the typical wider lobes of transition radiation for low velocities are common for both kinds of materials. In chiral matter this occurs when the chirality is relatively weak and the ratio between the particle's frequency $\omega$ and the medium resonant frequency $\omega_{rm}$ is lesser than one. In the nonreciprocal case, this happens when the topological parameter $\Delta$ is of the order of $\alpha$. This can be compared through the Figs. 4 and 5 of Ref. \cite{Galyamin} and our Figs. \ref{PLOTS B1} and \ref{PLOTS B2} for the 3D topological insulator TlBiSe$_2$.

While both kinds of electric fields share similar linear contributions in the chiral and nonreciprocal quantities, we find that in the nonreciprocal case there are contributions of higher order than one in the topological parameter $\Delta$. These contributions are included in the electric field component along $\mathbf{e}_{\theta1}$ through the modified Fresnel coefficients $R_{TM,TM}^{12}(\theta_1,\Delta)$ and $T_{TM,TM}^{12}(\theta_1,\Delta)$ in Eq. (\ref{E reflex whole}). Regarding the radiation patterns, Galyamin \textit{et al.} \cite{Galyamin} show that if the ratio between the particle frequency $\omega$ and the medium resonance frequency $\omega_{rm}$ is greater than one, then the chiral contributions are of the same order of magnitude and comparable regardless of the particle's velocity and the strength of the chirality, as their Figs. 4 and 5 illustrate. By comparing with our Figs. \ref{PLOTS C1} and \ref{PLOTS C2}, this differs from our findings, where the contributions of the topological parameter $\Delta$ become significant only in the ultrarelativistic regime.

Remarkably, in the ultrarelativistic regime there are two lobes for the chiral matter emitted at different angles, when the medium has strong chirality as Fig. 5 of Ref. \cite{Galyamin} show. This could explain why the chiral contributions do not increase the standard contributions and both remain with the same order of magnitude. Nevertheless, for a nonreciprocal medium the transition radiation has only one lobe that is emitted at the same angle of the standard contribution resulting in the enhancement of the whole radiative response as our Figs. \ref{PLOTS C1} and \ref{PLOTS C2} for the magnetoelectric TbPO$_4$ illustrate. This constitutes the main difference between chiral transition radiation vs nonreciprocal.

Finally, we give our two last comments of this Section. The first one is devoted to remark that the radiation displayed in the upper hemisphere of Figs. \ref{PLOTS B1}\,-\ref{PLOTS C2}, and in the lower one of Figs. \ref{PLOTS D1} and \ref{PLOTS D2} lies in the backward direction of the particle because the charge is moving from right to left along the line ($0-\pi$) showed in Figs. \ref{PLOTS B1} and \ref{PLOTS C1}, or from left to right along the line ($\pi-0$) showed in Fig. \ref{PLOTS D1}. Here we have only focused on this backward radiation because the study of the forward one will lead to the question on how to separate the Vavilov-\v{C}erenkov radiation from the transition radiation emitted forwards by the particle in the corresponding medium after it crosses the interface. The second one is devoted to two important limit cases from Eq. (\ref{Angular distribution TR}): (i) $\varepsilon_1=\varepsilon_2$ and $\Delta=0$, which easily gives 0 meaning that for an homogeneous standard dielectric medium there is no transition radiation constituting another consistency check for our results.  (ii) $\varepsilon_1=\varepsilon_2$ and $\Delta\neq0$ leads to the following angular distribution of radiation
\begin{eqnarray}
\frac{ d^2 \mathcal{E}_1 }{ d\omega d\Omega } &=& \frac{ q^2 }{ 4\pi\varepsilon_0 } \sqrt{ \frac{ \varepsilon_1 }{ \mu_1 }} \frac{ v^2 }{ 4\pi^2 c^3 } \frac{ \sin^2\theta_1 }{ (1 - \beta^2 n_1^2 \cos^2\theta_1 )^2 } \nonumber\\
&& \times \frac{ 4\Delta^2\mu_1^2 }{ (4\varepsilon_1\mu_1^3 + \Delta^2 )^2 } \left[ \Delta^2 + 4\mu_1^2\beta^2n_1^4\cos^2\theta_1 \right] \;, \label{Angular distribution TR LC1} \nonumber\\
\end{eqnarray}
which remarkably shows that transition radiation occurs even if both media have the same permittivites and is of order $\Delta^2$. The polar and standard plots of the transition radiation have the same behavior as Figs.  \ref{PLOTS B1}\,-\ref{PLOTS C2}, but rescaled to order $\Delta^2$, thereby we will omit them. This result contrasts clearly with standard electrodynamics and reinforces the idea that the topological parameter mimics a permittivity as analyzed in Refs. \cite{Crosse-Fuchs-Buhmann,OJF-LFU} for other kind of radiations. After comparing the angular dependence of Eqs. (\ref{Angular distribution TR Ideal}) and (\ref{Angular distribution TR LC1}), we observe that the first term resembles an interface made up of an ideal conductor and an upper medium. Furthermore, we can simplify Eq. (\ref{Angular distribution TR LC1}) for an interface between vacuum and a nonmagnetic strong three-dimensional TI with $\varepsilon_1=\varepsilon_2=\mu_1=\mu_2=1$ and $\Delta\neq0$ resulting that
\begin{eqnarray}
\frac{ d^2 \mathcal{E}_1 }{ d\omega d\Omega } &=& \frac{ q^2 }{ 4\pi\varepsilon_0 } \frac{ v^2 }{ 4\pi^2 c^3 } \frac{ \sin^2\theta_1 }{ (1 - \beta^2\cos^2\theta_1 )^2 } \nonumber\\
&& \times \frac{ 4\Delta^2 }{ (4 + \Delta^2 )^2 } \left[ \Delta^2 + 4\beta^2\cos^2\theta_1 \right] \;, \label{Angular distribution TR LC2}
\end{eqnarray}
which shows interestingly that transition radiation will occur for an interface with the same permittivity and permeability as the vacuum. In this way, from Eqs. (\ref{Angular distribution TR LC1}) and (\ref{Angular distribution TR LC2}) we conclude that such transition radiation has a pure topological origin. These last results for velocities $v<c/n_1$ complement the analysis of reversed Vavilov-\v{C}erenkov radiation in strong three-dimensional TIs made in Ref. \cite{OJF-LFU-ORT}, because from Eq. (\ref{Angular distribution TR LC1}), as well as from Eq. (\ref{Angular distribution TR}), we can recover such radiation when $v>c/n_1$.

%
\section{Frequency distribution of the radiation: Ultra-relativistic case} \label{ULTRA}
In this section, we calculate the frequency distribution of the radiation generated by the particle through its infinite path along the $z$ axis, which is obtained by integrating the angular distribution (\ref{Angular distribution TR}) of radiation over the solid angle $\Omega$
\begin{equation} \label{Freq distribution}
\frac{ d \mathcal{E} }{ d\omega } = \int_\Omega \frac{ d^2 \mathcal{E} }{ d\omega d\Omega } d\Omega \;.
\end{equation}

Unfortunately the general case of Eq. (\ref{Freq distribution}) cannot be integrated over the solid angle $\Omega$ analytically, but numerically can be performed once all the parameters are specified. However, we can provide two analytical results if we restrict ourselves to two cases. The first one will be when the particle moves from vacuum with $\varepsilon_1=\mu_1=1$ and $\Theta_1=0$ into a nonmagnetical strong three-dimensional TI with $\mu_2=1$, which is of practical interest for particle detectors. The second one will be the opposite situation, when the particle moves from a non-magnetical strong three-dimensional TI with $\mu_2=1$ into vacuum. Additionally, our aim will require to study both cases only in the ultra-relativistic regime. 

Let us begin with the vacuum-to-TI case. It is well-known that the radiation of an ultra-relativistic particle with mass $m$ and total energy $E$ has a sharp maximum in the direction $\theta_1\sim mc^2/E \sim \sqrt{1-\beta^2}$ \cite{Garibian, Landau-Lifshitz, Jackson}, as can be appreciated in Figs. \ref{PLOTS C1} and \ref{PLOTS C2}, which means that the maximum appears at small values of $\theta_1$.  In this way, to carry out the integration indicated in Eq. (\ref{Freq distribution}), it is convenient to evaluate all the terms with a weak angular dependence under the integral sign by setting $\theta_1\sim 0$ in these expressions \cite{Garibian}. Then, Eq. (\ref{Angular distribution TR}) becomes
\begin{equation}
\frac{ d^2 \mathcal{E}_1 }{ d\omega d\Omega } = \frac{ q^2 }{ 4\pi\varepsilon_0c }  \frac{ \mathcal{F}(\varepsilon_2,\Delta) (1+\sqrt{\varepsilon_2} )^2 \theta_1^2 }{ \pi^2(1 - \beta^2 + \theta^2_1)^2 \left( 1 + \beta \sqrt{\varepsilon_2 -\theta_1^2} \right)^2 } \;, \label{ADTR VTI 1}
\end{equation}
where we defined
\begin{eqnarray}
\mathcal{F}(\varepsilon_2,\Delta) & = & \left[ R_{TM,TM}^{12}(0,\Delta) \right]^2 + \left[ R_{TE,TM}^{12}(0,\Delta) \right]^2 \;, \nonumber\\
& = & \frac{ [ (\varepsilon_2 - 1) + \Delta^2 ]^2 +  4\Delta^2 }{ [ ( \sqrt{\varepsilon_2} + 1 )^2 + \Delta^2 ]^2 } \;,
\end{eqnarray}
by evaluating the modified Fresnel coefficients at $\theta_1=0$, which can be regarded as their retarded limit \cite{Fuchs-Crosse-Buhmann}.

If $\varepsilon_2$ is not too close to unity, we can replace the last factor in the denominator of Eq. (\ref{ADTR VTI 1}) by $(1+ \sqrt{\varepsilon_2} )^2$ \cite{Landau-Lifshitz}. After considering this last approximation and substituting in Eq. (\ref{Freq distribution}), it gives the following frequency distribution of the radiation, with logarithmic accuracy:
\begin{eqnarray}
\frac{ d \mathcal{E}_1 }{ d\omega } &\simeq& \int_0^{\sim 1} \frac{ d^2 \mathcal{E}_1 }{ d\omega d\Omega }\; 2\pi\theta_1 d\theta_1 \;, \nonumber\\
&=& \frac{ q^2 }{ 4\pi\varepsilon_0c }  \frac{ \mathcal{F}(\varepsilon_2,\Delta) }{ \pi }  \left[ \frac{ 1 }{ \beta^2 - 2} + \ln\left( \frac{ 2 - \beta^2 }{ 1 -  \beta^2 } \right) \right] \;, \nonumber\\
 & \simeq & \frac{ q^2 }{ 4\pi\varepsilon_0c }  \frac{ \mathcal{F}(\varepsilon_2,\Delta) }{ \pi } \ln \left( \frac{ 1 }{ 1 -  \beta^2 } \right) \;. \label{Freq distribution VTI}
\end{eqnarray}
This expression shows a dependence on the topological parameter $\Delta$ meaning that in this ultra-relativistic regime the frequency distribution of the radiation is sensible to changes of this parameter, as can be seen at the level of the angular distribution in Figs. \ref{PLOTS C1} and \ref{PLOTS C2}. 

It is worth to analyze the frequency distribution within the perspective of the limit cases discussed at the end of Sec. \ref{ANGULAR}. First, for $\varepsilon_1=\varepsilon_2=1$ and $\Delta=0$ we found a zero angular distribution, so its frequency distribution is zero too. Secondly, if $\varepsilon_1=\varepsilon_2=1$ and $\Delta\neq0$ led to the angular distribution of Eq. (\ref{Angular distribution TR LC2}). After applying the same ultrarelativistic approximation used above, Eq. (\ref{Angular distribution TR LC2}) is rewritten as 
\begin{equation}
\frac{ d^2 \mathcal{E}_1 }{ d\omega d\Omega } = \frac{ q^2 }{ 4\pi\varepsilon_0c } \frac{ \Delta^2 }{ \pi^2 (4 + \Delta^2 ) } \frac{  \theta_1^2 }{ (1 - \beta^2 + \theta^2_1)^2 }\;,
\end{equation}
which clearly is of order $\Delta^2$. Then, along the same lines of Eq. (\ref{Freq distribution VTI}) we compute its frequency distribution leading to
\begin{equation}
\frac{ d \mathcal{E}_1 }{ d\omega } \simeq \frac{ q^2 }{ 4\pi\varepsilon_0c } \frac{ \Delta^2 }{ \pi (4 + \Delta^2 ) } \ln \left( \frac{ 1 }{ 1 -  \beta^2 } \right) \;,
\end{equation}
which again shows a nonzero result when the strong three-dimensional TI and the vacuum have the same permittivity and permeability. 

Lastly, we consider the TI-to-vacuum case. As discussed at the end of Sec. \ref{ANGULAR}, this problem differs from the former one only by a change in the sign of the velocity $v$ and the indexes 1 to 2 should be reversed in Eq. (\ref{Angular distribution TR}). Thus we apply the same ultrarelativistic approximation done for Eq. (\ref{ADTR VTI 1}) finding that 
\begin{equation}\label{ADTR TIV 1}
\frac{ d^2 \mathcal{E}_2 }{ d\omega d\Omega } = \frac{ q^2 }{ 4\pi\varepsilon_0c }  \frac{ \sqrt{ \varepsilon_2 } (1 - \beta )^2 \theta_2^2 }{ \pi^2 (1 - \beta^2 \varepsilon_2 + \theta^2_2)^2 \left( 1 - \beta \sqrt{1 - \varepsilon_2 \theta_2^2} \right)^2 }  \;.
\end{equation}
Again, we have to impose an additional condition in order to carry out the integration over the solid angle defined by the angle $\theta_2$. So, as long as $\varepsilon_2 \theta_2^2 \ll 1$ we can approximate the last factor in the denominator of Eq. (\ref{ADTR TIV 1}) by $1-\beta$.  After inserting this last approximation into Eq. (\ref{ADTR TIV 1}) and repeating the same angular integral of Eq. (\ref{Freq distribution VTI}) results the next frequency distribution of radiation, with logarithmic accuracy,
\begin{equation}\label{Freq distribution TIV}
\frac{ d \mathcal{E}_2 }{ d\omega } \simeq  \frac{ q^2 }{ 4\pi\varepsilon_0c }  \frac{ \sqrt{ \varepsilon_2 } }{ \pi } \ln \left( \frac{ 1 }{ 1 -  \beta^2 \varepsilon_2 } \right) \;. 
\end{equation}

Remarkably, this distribution is independent on the topological parameter $\Delta$ being completely the opposite situation of the one given by Eq. (\ref{Freq distribution VTI}), meaning that the standard transition radiation is the dominant phenomenon for the TI-to-vacuum case. This result accounts for the directivity of this phenomenon discussed at the end of Sec. \ref{ANGULAR}, but now at the level of the frequency distribution.

Finally, from Eqs. (\ref{Freq distribution VTI}) and (\ref{Freq distribution TIV}) we observe that they have the same logarithmic dependence on the velocity as the frequency distribution of total energy for standard dielectric media obtained by Garibian \cite{Garibian, Landau-Lifshitz}. Nevertheless, there are important differences to point out between the Garibian's ones and the quantities reported in Eqs. (\ref{Freq distribution VTI}) and (\ref{Freq distribution TIV}), which we describe briefly. We start with the Garibian's expression for Eq. (\ref{Freq distribution VTI}), which only works for ordinary dielectric media. So, to connect both expressions, $\Delta=0$ must be taken. Then, $\mathcal{F}(\varepsilon_2,\Delta=0)$ should be replaced by the coefficient $ ( \sqrt{ \varepsilon_2 } - 1)^2 / ( \sqrt{ \varepsilon_2 } + 1 )^2$ \cite{Garibian, Landau-Lifshitz}. Then, let us move on to Garibian's expression for Eq. (\ref{Freq distribution TIV}), which only works for ordinary dielectric media as well as ours. Although the connection between both quantities is achieved by omitting the factor $\sqrt{ \varepsilon_2 }$, our frequency distribution works only for $v\in[0,c/\sqrt{ \varepsilon_2 })$ and not in the full range  $v\in[0,c)$ as Garibian found \cite{Garibian, Landau-Lifshitz}. The origin of these discrepancies comes from the fact that both quantities although similar they are not the same, therefore it is not permissible to compare them. Garibian's version of Eq. (\ref{Angular distribution TR}) describes the angular distribution of the total energy of the radiation and in consequence his version of Eqs. (\ref{Freq distribution VTI}) and (\ref{Freq distribution TIV}) describe the frequency distribution of the total energy of the radiation \cite{Frank Lecture}, which are obtained through a different procedure called Hamiltonian method described in Refs. \cite{Landau-Lifshitz, Ginzburg Book} after decoupling the normal modes of electrodynamics. In contrast, ours correspond to the angular distribution of the radiation and its frequency distribution. Here we followed the approach exposed in Refs. \cite{Frank-Ginzburg, Frank Lecture} and used Garibian's expressions only as a guide, because the decoupling of the normal modes in the modified Maxwell equations (\ref{Gauss B})-(\ref{Ampere-Maxwell}) is tricky to perform due to the Dirac delta coming from the gradient of the axion coupling given by Eq. (\ref{THETA}).

\section{Conclusions} \label{CONCLUSIONS}
We have analyzed the transition radiation produced by a charged particle propagating with constant velocity $v$ along the $z$ axis that crosses the interface between two generic magnetoelectric media with special emphasis on TIs. To achieve this goal, we employed the Green's function of two layered three-dimensional TIs with different permittivities, permeabilities and topological parameters in order to obtain the electromagnetic field in terms of Hankel transforms. By means of the far-field approximation together with the steepest descent method, we were able to obtain analytical expressions for the electromagnetic field. As a consequence of this approach, the resulting field is a superposition of spherical waves and lateral waves with contributions of both kind associated to a purely topological origin. The calculation of the angular distribution of the radiation shows that in a region far from the interface the main contribution is due to the spherical waves.  In such a region of space, the main characteristics of the transition radiation modified by strong three-dimensional TIs are the following. (i) The angles of maximal emission are the same as in the standard case. (ii) The directivity of the phenomenon remains as in the standard case, which means that the radiation patterns depend on the sign particle velocity. Remarkably, the additional contributions from the strong three-dimensional TI are more evident when the particle parts from the vacuum, crosses the interface and continues its path through the TI. (iii) As in standard electrodynamics the transition radiation and Vavilov-\v{C}erenkov radiation can coexist, unless the charge velocity $v$ is lower than the speed of light in both media assuring that transition radiation will occur alone in the backwards direction relative to the particle movement.  

After ruling out the arising of Vavilov-\v{C}erenkov radiation, we studied two important configurations: vacuum-to-TI case and the TI-to-vacuum one. To understand both cases we chose two materials: the topological insulator TlBiSe$_2$ with $\varepsilon_2=4$, $\mu_2=1$, and $\Delta=11\alpha$, and to enhance the new effects the magnetoelectric TbPO$_4$ with $\varepsilon_2=3.4969$, $\mu_2=1$, and $\Delta=0.22$. The results of the vacuum-to-TI case when the $v=0.75c$ and the material is the topological insulator TlBiSe$_2$ show that the new contributions are almost imperceptible as Figs. \ref{PLOTS B1} and \ref{PLOTS B2} illustrate. We find numerically that the differences are of the order of $10^{-3}$ for $\Delta=11\alpha$ being the highest value for such strong three-dimensional TI. The differences decreases for lower values of $\Delta$, i.e., the modifications to the transition radiation become more insignificant. However, the results in the ultrarelativistic regime for $v=0.99c$ and the magnetoelectric TbPO$_4$ are more interesting because the corresponding maximum is greater than the one related to standard transition radiation. This means that the presence of the strong three-dimensional TI increases the magnitude of the backward radiation considerably in the ultrarelativistic regime as can be appreciated in Figs. \ref{PLOTS C1} and \ref{PLOTS C2}. Regarding the TI-to-vacuum case, we used again the magnetoelectric TbPO$_4$ as material but even in the ultrarelativistic regime the additional contributions are not significant because they are dominated by the standard transition radiation, as we presented in Figs. \ref{PLOTS D1} and \ref{PLOTS D2}. We also analyzed two important limit cases. The first one happens when both magnetoelectrics become a single standard homogeneous dielectric medium, which is described by the parameters $\varepsilon_1=\varepsilon_2$ and $\Delta=0$. For this case, we found null transition radiation as in standard electrodynamics. The second limit case resulted more interesting because both magnetoelectrics have the same optical properties but different topological parameters allowing the emergence of new effects. For such case, we obtained that transition radiation occurs and is of order $\Delta^2$. This situation contrasts with standard electrodynamics where necessary $\varepsilon_1\neq\varepsilon_2$ and shows that the topological parameter mimics the role of the permittivity. Moreover, this limit case and its pure topological transition radiation are reminiscent of what is reported in the work \cite{AMR-LFU}, where the new optical effects are enhanced when $\varepsilon_1=\varepsilon_2$ and $\mu_1=\mu_2=1$ when a static hydrogenlike ion interacts with a planar  three-dimensional TI. Although this limit case results interesting, in practice realizing the same permittivities for the strong three-dimensional TI and the other dielectric in a certain setup could be complicated to achieve. Therefore our proposal with two different permittivities could be a bit easier to build in laboratory.
 
In this work we investigated the frequency distribution of this transition radiation too. Notwithstanding that a general expression for the frequency distribution cannot be provided analytically, we were able to integrate the angular distribution of radiation by restricting ourselves to the ultrarelativistic regime and considering a nonmagnetic strong three-dimensional TI. For the vacuum-to-TI case, we found that its frequency distribution depends logarithmically on the velocity and is proportional to the sum of two squared modified Fresnel coefficients, namely $\left[ R_{TM,TM}^{12}(\theta_1= 0,\Delta) \right]^2 + \left[ R_{TE,TM}^{12}(\theta_1=0,\Delta) \right]^2$. In this framework, we also studied the frequency distribution of the pure topological transition radiation arisen in the limit case when the strong three-dimensional TI has the same optical properties of the vacuum. For such case, we found that the corresponding frequency distribution also depends logarithmically on the velocity and is of the order $\Delta^2$. On the other hand, we obtained the frequency distribution for the TI-to-vacuum case also depends logarithmically on the velocity but it is independent on the topological parameter, which explains why the angular distribution of this case is unaffected by the presence of the strong three-dimensional TI as Figs. \ref{PLOTS D1} and \ref{PLOTS D2} show. 

In order to observe our results in the ultrarelativistc regime, concretely those shown in Figs. \ref{PLOTS C1} and \ref{PLOTS C2}, the setup should contain a source of impinging charges, a radiator (the 3D TI or a magnetoelectric) and a transition radiation detector. First, we remark that the particle's velocity is independent of any internal characteristic of the magnetoelectric or TI and this radiation is measured by detectors located backwards to the particle's trajectory. Second, for an interface vacuum-TI or a vacuum-magnetoelectric the source should provide a spatially localized group of electrically charged particles, that have approximately the same trajectory, kinetic energy and direction. The required condition on the velocity for the particles velocity is dictated by Eq. (\ref{v choice}). Given that the source should be located in vacuum,  the particles velocity can be $v=0.99c$ or even greater. Next, if the single layer radiator is selected as a 3D topological insulator, it would be advisable to break the time-reversal symmetry by doping the surfaces with thin ferromagnetic films instead of switching on an external field, since we have assumed that charged particles are moving with constant velocity coming from vacuum and crossing the material. Though transition radiation can be detected for a single layer radiator, in practice optimization of particle discrimination leads to realize multi-foil radiators constituted by a certain number of foils of thickness $l_1$ separated by a medium (usually a gas) of thickness $l_2$ \cite{Andronic}. In this way, our present work constitutes the first step on the study of such possible configurations, which we leave as future work. Then, according to Figs. \ref{PLOTS C1} and \ref{PLOTS C2}, for ultrarelativistic velocities, the emission angle of transition radiation is closer to $0$ measured from the particle's trajectory meaning that the transition radiation detector must be situated closer to the beam. Lastly, the detection of the present transition radiation could be made with current state-of-the-art transition radiation detectors because our results provide an enhancement to the standard transition radiation.

Finally, we remark that the pure topological transition radiation and the transition radiation enhancement in the ultrarelativistic regime for the vacuum-to-TI case offer an indirect measurement of the topological parameter $\Delta$ as some works in the literature have proposed \cite{Wu, Dziom, Okada}. Furthermore, this work can be relevant for the current research of dark matter detection where axions are a well-motivated candidate \cite{Sikivie}, because for such detection condensed matter physics offers new tools  through TIs \cite{Nenno et al}, antiferromagnetically doped TIs \cite{Marsh et al, Jan Schuette} or multiferroics \cite{Roising et al} by exploiting the properties of the magnetoelectric effect also responsible of our results.

\section{Acknowledgments}
O. J. F. has been supported by the postdoctoral fellowships CONACYT-770691 and CONACYT-800966.

\appendix

\section{Green's Function components relevant for the electric field} \label{A}
To the purpose of this article be self-contained in this appendix we provide the necessary details of the Green's function $\mathbb{G}(\mathbf{r},\mathbf{r}';\omega)$ required to obtain the electric field via Eq. (\ref{E GF j}) for a source and field point in an arbitrary layer. First, the free-space part $\mathbb{G}^{(0)}(\mathbf{r},\mathbf{r}';\omega)$ of the Green's function, proportional to the unit tensor $\boldsymbol{\mathit{I}}$, is given by
\begin{eqnarray} \label{GF free}
\mathbb{G}^{(0)}(\mathbf{r},\mathbf{r}';\omega) &=& \mu(\omega) \frac{ e^{ \mathrm{i} kR } }{ 4\pi R } \left[ \left( 1 + \frac{ \mathrm{i}kR - 1 }{ k^2 R^2 } \right) \boldsymbol{\mathit{I}} \right. \nonumber\\
&& \left. + \frac{ 3 - 3\mathrm{i}kR - k^2R^2 }{ k^2 R^2 } \frac{ \mathbf{R} \otimes \mathbf{R} }{ R^2 } \right] \;,
\end{eqnarray}
where $\mathbf{R} = \mathbf{r} - \mathbf{r}'$.

Then the reflective part of the Green's function $\mathbb{G}^{(1)}$ of Eq. (\ref{GF split}) can be rewritten as 
\begin{equation} \label{GF reflex}
\mathbb{G}^{(1)\; ij}(\mathbf{r},\mathbf{r}';\omega)=\int \frac{ d^2\mathbf{k}_\parallel }{ (2\pi)^2 } e^{ \mathrm{i}\mathbf{k}_\parallel\cdot\mathbf{R}_\parallel } R^{ ij } \left( z,z';\mathbf{k}_\parallel,\omega \right) \;,
\end{equation}
with $\mathbf{R}_\parallel=(x-x',y-y')$, $\mathbf{k}_\parallel=(k_x,k_y)$ is the transversal wavevector (parallel to the interface), and whose only required components in the present work are
\begin{eqnarray}
&& R^{\, xz } \left( z,z';\mathbf{k}_\parallel,\omega \right) =  \frac{ \mathrm{i} \mu(\omega) }{ 2k_{z} }e^{ \mathrm{i}k_{z}(|z|+|z'|) } \nonumber\\
&& \times \left[ -\mathrm{sgn}(z') \frac{ k_x k_{z} }{ k^2 } R_{TM,TM} + \frac{ k_y }{ k }R_{TE,TM} \right] ,
\end{eqnarray}
\begin{eqnarray}
&& R^{\, yz } \left( z,z';\mathbf{k}_\parallel,\omega \right) = \frac{ \mathrm{i} \mu(\omega) }{ 2k_{z} }e^{ \mathrm{i}k_{z}(|z|+|z'|) } \nonumber\\
&& \left[-\mathrm{sgn}(z') \frac{ k_y k_{z} }{ k^2 } R_{TM,TM} - \frac{ k_x }{ k } R_{TE,TM} \right] \;,
\end{eqnarray}
\begin{equation}
R^{\, zz } \left( z,z';\mathbf{k}_\parallel,\omega \right) = \frac{ \mathrm{i} \mu(\omega) }{ 2k_{z} }e^{ \mathrm{i}k_{z}(|z|+|z'|) } \left[ \frac{ k_\parallel^2 }{ k^2 } R_{TM,TM} \right] \;,
\end{equation}
where $\mu_1(\omega)$ is the permeability of the upper medium, $\mathrm{sgn}$ stands for the sign function and we recall that $k_{z}=\sqrt{k^2-\mathbf{k}_\parallel^2}$. 

Similarly, the transmissive part of the Green's function $\mathbb{G}^{(1)}$ of Eq. (\ref{GF split}) can be rewritten as 
\begin{equation}  \label{GF trans}
\mathbb{G}^{(1)\; ij }(\mathbf{r},\mathbf{r}';\omega)=\int \frac{ d^2\mathbf{k}_\parallel }{ (2\pi)^2 } e^{ \mathrm{i}\mathbf{k}_\parallel\cdot\mathbf{R}_\parallel } T^{ ij } \left( z,z';\mathbf{k}_\parallel,\omega \right) \;,
\end{equation}
whose only required components in the present work are 
\begin{eqnarray}
&& T^{\, xz } \left( z,z';\mathbf{k}_\parallel,\omega \right) = \frac{ \mathrm{i} \mu^{ \, \prime } (\omega) }{ 2k_{z'} } e^{\mathrm{i}k_{z}|z|+\mathrm{i}k_{z'}|z'|} \nonumber\\
&& \times \left[ \mathrm{sgn}(z') \frac{ k_x k_{z} }{ k k' } T_{TM,TM} + \frac{ k_y }{ k } T_{TE,TM} \right] \;, 
\end{eqnarray}
\begin{eqnarray}
&& T^{\, yz } \left( z,z';\mathbf{k}_\parallel,\omega \right) = \frac{ \mathrm{i} \mu^{ \, \prime } (\omega) }{ 2k_{z'} } e^{\mathrm{i}k_{z}|z|+\mathrm{i}k_{z'}|z'|} \nonumber\\
&& \times \left[ \mathrm{sgn}(z') \frac{ k_y k_{z} }{ k k' } T_{TM,TM} - \frac{ k_x }{ k' } T_{TE,TM} \right] \;, 
\end{eqnarray}
\begin{eqnarray}
&&T^{\, zz } \left( z,z';\mathbf{k}_\parallel,\omega \right) = \frac{ \mathrm{i} \mu^{ \, \prime } (\omega) }{ 2k_{z'} } e^{\mathrm{i}k_{z}|z|+\mathrm{i}k_{z'}|z'|} \nonumber\\
&& \times \left[ \frac{ k_\parallel^2 }{ k k' } T_{TM,TM} \right] \;,
\end{eqnarray}
where $k_{z'}=\sqrt{k^{' \, 2} - \mathbf{k}_\parallel^2}$,  $k'$ and $\mu^{ \, \prime }(\omega)$ denote the wavenumber and the permeability in the source medium. 

Once the labels for the upper and lower media are designed. Through Eqs. (\ref{GF free}), (\ref{GF reflex}) and (\ref{GF trans}) we obtain the four possible cases for a two layer configuration:  observer and source at the upper medium, observer and source at the lower medium, observer at the upper medium and source at the lower one, and vice versa.

\section{Approximating the required integrals for the electric field} \label{B}
In this appendix, we apply the steepest descent method \cite{Banhos,Chew,Mandel, Chew art, Wait, OJF-LFU} to find the far-field approximation of the $k_\parallel$ integrals of the reflective (\ref{Ex1 int})-(\ref{Ez1 int}) and transmissive electric fields (\ref{Ex2 int})-(\ref{Ez2 int}). Nevertheless, we will only present the calculations for the $x$ component given by Eq. (\ref{Ex1 int}) in full detail. The other relevant components are obtained in an analogous fashion. To this end, we closely follow  Ref. \cite{Banhos}, which provides  a detailed account of the general procedure to deal with the steepest descent method for these class of integrals.

The two $k_\parallel$ integrals defined in Eqs. (\ref{integral I1 napp}) and (\ref{integral I2 napp}) that need to be computed in Eq.(\ref{Ex1 int}) are
\begin{equation}
\mathcal{I}_1  =  \int_0^\infty k_\parallel dk_\parallel R_{TM,TM}^{12}(k_\parallel) J_0(k_\parallel R_\parallel) e^{\mathrm{i}k_{z,1}(|z|+|z'|)} \;, \label{integral I1} 
\end{equation}
\begin{equation}
\mathcal{I}_2  =  \int_0^\infty \frac{ k_\parallel dk_\parallel }{ k_{z,1} } R_{TE,TM}^{12}(k_\parallel) J_0(k_\parallel R_\parallel) e^{\mathrm{i}k_{z,1}(|z|+|z'|)} \;, \label{integral I2} 
\end{equation}
where $R_{TM,TM}^{12}$ and $R_{TE,TM}^{12}$ are given by Eqs. (\ref{Fresnel 1}) and (\ref{Fresnel 2}).
\begin{figure}[tbp]
\includegraphics[width=0.45\textwidth]{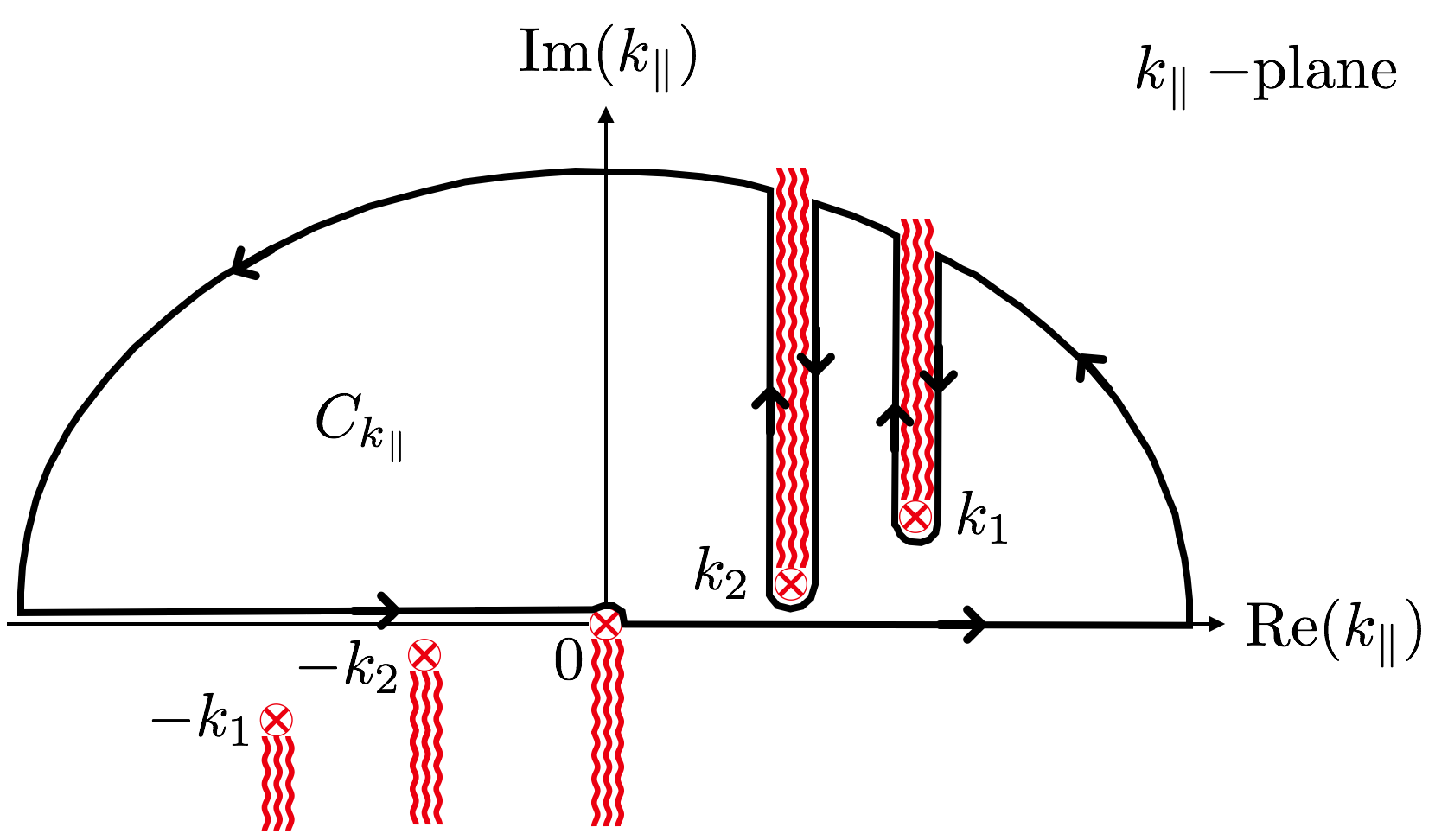}
\caption{ The Sommerfeld path of integration $C_{k_{\parallel}}$ in the $k_{\parallel}$-plane showing the branching cuts originating the branching points $\pm k_1$, $\pm k_2$ and 0.  }
\label{SOMMERFELD}
\end{figure}

Let us begin with $\mathcal{I}_1$. First, we go to the complex plane by writing the Bessel function in terms of the Hankel function $H^{(1)}_0(k_\parallel R_\parallel)$ and employing the reflection formula \cite{Arfken}, which allows us to extend the integration interval to $-\infty$. The result is 
\begin{equation}
\mathcal{I}_1 = \frac{1}{2} \oint_{C_{k_{\parallel}}} k_\parallel dk_\parallel R_{TM,TM}^{12}(k_\parallel) H^{(1)}_0(k_\parallel R_\parallel) e^{\mathrm{i}k_{z,1}(|z|+|z'|)} \;,
\end{equation}
where $C_{k_{\parallel}}$ is the Sommerfeld path of integration illustrated in Fig. \ref{SOMMERFELD}. This path avoids the branch cuts dictated by the Hankel function $H^{(1)}_0$ at the origin, $k_{z,1}=\sqrt{k_1^2-\mathbf{k}_\parallel^2}$ and $k_{z,2}=\sqrt{k_2^2-\mathbf{k}_\parallel^2}$ due to the exponential and $R_{TM,TM}^{12}$. Because we have considered an absorbing media with $\mathrm{Im}(n_1),\mathrm{Im}(n_2)>0$ the exponential will be negative assuring the required convergence \cite{Banhos}. However, as the observer's zenith angle $\theta_1$ increases towards $\pi/2$ the path $C_{k_{\parallel}}$ eventually will pass through the branch point at $k_\parallel=k_2$ for the discarding angle $\theta_1^{disc}$ given by Eq. (\ref{theta UH}). For $\theta_1\in(\theta_1^{disc},\pi/2]$ crosses the branch cut for $k_{z,2}$ and changes to another sheet of the Riemann surface of the squared root. Thus, in order to return to the starting sheet we choose the deformation of the path of integration described and justified by Ba\~{n}os \cite{Banhos}. Consequently, if we denote by $\mathcal{J}_1$ and $\mathcal{J}_2$ the evaluations of the integral $\mathcal{I}_1$ along such deformation path, we may write 
\begin{equation} \label{Integrals I and Js}
\mathcal{I}_1 = \mathcal{J}_1 +H(\theta_1-\theta_1^{disc})\mathcal{J}_2\;,
\end{equation}
where
\begin{equation}
\mathcal{J}_1  =  \frac{1}{2}\int_{C_{1}} k_\parallel dk_\parallel R_{TM,TM}^{12}(k_\parallel) H^{(1)}_0(k_\parallel R_\parallel) e^{\mathrm{i}k_{z,1}(|z|+|z'|)} \;, \label{integral J1} 
\end{equation}
\begin{equation}
\mathcal{J}_2  =  \frac{1}{2}\int_{C_{2}} k_\parallel dk_\parallel R_{TM,TM}^{12}(k_\parallel) H^{(1)}_0(k_\parallel R_\parallel) e^{\mathrm{i}k_{z,1}(|z|+|z'|)} \;, \label{integral J2}
\end{equation}
with $C_{1}$ as the path without crossing the branch point for $\theta_1\in[0,\theta_1^{disc})$ and $C_{2}$ is the path other case for $\theta_1\in[\theta_1^{disc},\pi/2]$.

Let us compute first $\mathcal{J}_1$. We begin by applying the conformal transformation $k_\parallel=k_1\sin\gamma$ obtaining
\begin{eqnarray}
&& \mathcal{J}_1  =  \frac{k_1^2}{2} \int_{C_{1\gamma}} d\gamma \sin\gamma\cos\gamma R_{TM,TM}^{12}(\gamma) \nonumber\\
&& \times H^{(1)}_0(k_1\sin\gamma R_\parallel) e^{\mathrm{i}k_1\cos\gamma(|z|+|z'|)} \;.
\end{eqnarray}
Then, we employ the asymptotic expansion of the Hankel function \cite{Abramowitz} because we are interested only in the far-field regime. In this way, we have
\begin{eqnarray} 
&& \mathcal{J}_1  =  e^{-\mathrm{i}\pi/4} k_1^2 \sqrt{ \frac{ 1 }{ 2\pi k_1 R_\parallel} } \oint_{C_{1\gamma}} d\gamma \sqrt{\sin\gamma} \cos\gamma  \nonumber\\
&& \times R_{TM,TM}^{12}(\gamma) e^{\mathrm{i} k_1R_\parallel \sin\gamma + \mathrm{i} k_1\cos\gamma(|z|+|z'|)} .
\end{eqnarray}
Due to the location of the observer and the semi-infinite path of the particle at the upper hemisphere, $|z|=z$ and $|z'|=z'$. So, we write $|z|=r\cos\theta_1$ and $R_\parallel=r\sin\theta_1$, i.e., $r=\sqrt{R_\parallel^2+z^2}$. Thereby, we find 
\begin{eqnarray}   
\mathcal{J}_1 & = & e^{-\mathrm{i}\pi/4} k_1^2 \sqrt{ \frac{ 1 }{ 2\pi k_1 r\sin\theta_1} } \int_{C_{1\gamma}} d\gamma \sqrt{\sin\gamma} \cos\gamma \nonumber\\
&& \times R_{TM,TM}^{12}(\gamma) e^{\mathrm{i} k_1 r\cos\left(\gamma-\theta_1\right) + \mathrm{i}k_1z'\cos\gamma}\;.
\end{eqnarray}
Next we determine the saddle point of $\mathcal{J}_1$ by choosing the stationary phase as only  $\varphi\left(\gamma\right) = \mathrm{i}k_1r\cos\left(\gamma-\theta_1\right)$, according to Ref. \cite{Banhos}. Through $\varphi^{\prime} \left(\gamma_s\right)=0$ results that $\gamma_s=\theta_1$ is the saddle point, which leads to the full stationary phase $\mathrm{i}k_1r  \cos \theta_1$.  At this stage, the steepest descent path is specified on the $\gamma$ plane by demanding the next condition $\mathrm{Im} \left[ \varphi\left(\gamma\right) \right] = \mathrm{Im} \left[ \varphi\left(\gamma_s\right) \right]$ implying that $\mathrm{Im} \left[ \mathrm{i}k_1r\cos\left(\gamma-\theta_1\right) \right] = \mathrm{Im} \left[ \mathrm{i}k_1r \right]$ over $C_{1\gamma}$. Now, we shift the origin to coincide with the saddle point  by setting  $w=\gamma-\theta_1$ in $\mathcal{J}_1$,  which yields 
\begin{eqnarray} 
&& \mathcal{J}_1  =  e^{-\mathrm{i}\pi/4} k_1^2 \sqrt{ \frac{ 1 }{ 2\pi k_1 r\sin\theta_1} } \int_{C_{1w}} dw \sqrt{\sin(w + \theta_1)} \nonumber\\ 
&& \times \cos(w + \theta_1) R_{TM,TM}^{12}(w + \theta_1) e^{\mathrm{i} k_1[ r\cos w + z'\cos(w + \theta_1) ] }\;, \nonumber\\
\end{eqnarray}
where the reparametrized path  $C_{1w}$ satisfies $\mathrm{Im}\left[\mathrm{i}k_1r\cos w\right]=\mathrm{Im}\left[\mathrm{i}k_1r\right]$.

The following step is to introduce the conformal transformation $u^2/2=\varphi\left(0\right)-\varphi\left(w\right)=\mathrm{i}k_1r\left(1-\cos w\right)$, whose purpose is to map the steepest descent path into the real axis. This requires the change of variable $\cos w=1-{u^2}/{2\mathrm{i}k_1r}$ in $\mathcal{J}_1$, after which we obtain
\begin{equation}
\mathcal{J}_1 = \frac{k_1}{ \mathrm{i} }\sqrt{\frac{1}{2\pi\sin\theta_1}}\frac{e^{\mathrm{i}k_1r}}{r}\int_{C_{1u}}duF_1(u)e^{-u^2/2}\;,
\end{equation}
with
\begin{eqnarray} 
&& F_1(u)  =  \sin^{1/2}\left[w(u) + \theta_1\right] \cos\left[w(u) + \theta_1\right] \nonumber\\
&& \times R_{TM,TM}^{12} \left[w(u) + \theta_1\right] \frac{ e^{\mathrm{i}k_1z'\cos\left[w(u) + \theta_1\right]} }{ \sqrt{1-\frac{u^2}{4\mathrm{i}k_1r}} } \;. 
\end{eqnarray}
Then, we look at the behavior of $F_1(u)$ and find that it has only one branching point when the squared root vanishes. Here we will neglect such branch point because it matters only if we seek corrections of higher order than $r^{-1}$. As mentioned, the $u$ transformation mapped the steepest descent path into the real axis. Hence, we proceed to approximate $\mathcal{J}_1$ with standard calculus techniques. Since we are interested only in the dominant term of $\mathcal{J}_1$, it is enough to consider the zeroth-order term in the expansion of  $F_1(u)$ in its Taylor series around $u=0$ ($w=0$), because most of the contribution arises from its  vicinity due to the presence of the Gaussian function $e^{-u^2/2}$. Performing this, we obtain
\begin{equation} \label{final integral J1} 
\mathcal{J}_1 = \frac{ k_1 }{ \mathrm{i} } \frac{ e^{\mathrm{i}k_1r} }{ r } \cos\theta_1 R_{TM,TM}^{12} (\theta_1)e^{ \mathrm{i}k_1z'\cos\theta_1 } \;.
\end{equation}

Now, we move on to compute the integral $\mathcal{J}_2$ defined in Eq. (\ref{integral J2}), which captures the contribution of the branch point. We start by applying the conformal transformation $k_\parallel=k_2\cos\eta$ obtaining
\begin{eqnarray}    
&& \mathcal{J}_2  =  \frac{k_2^2}{2} \int_{C_{2\eta}} d\eta \cos\eta\sin\eta R_{TM,TM}^{12}(\eta) \nonumber\\
&& \times H^{(1)}_0(k_2\cos\eta R_\parallel) e^{\mathrm{i}\sqrt{ k_1^2 - k_2^2 \cos^2\eta } (|z|+|z'|)} \;. 
\end{eqnarray}
Then, we expand the Hankel function asymptotically due to our interest in the far-field regime. Hence, we have
\begin{eqnarray}  %
&& \mathcal{J}_2 = e^{-\mathrm{i}\pi/4} k_2^2 \sqrt{ \frac{ 1 }{ 2\pi k_2 R_\parallel} } \oint_{C_{2\eta}} d\eta \sqrt{\cos\eta}  \sin\eta \nonumber\\
&& \times R_{TM,TM}^{12}(\eta) e^{\mathrm{i} k_2R_\parallel \cos\eta + \mathrm{i} \sqrt{ k_1^2 - k_2^2 \cos^2\eta } (|z|+|z'|)}\;. \nonumber\\
\end{eqnarray} 
Because the observer and the semi-infinite path of the particle are both located at the upper hemisphere, $|z|=z$ and $|z'|=z'$. So, we write $|z|=r\cos\theta_1$ and $R_\parallel=r\sin\theta_1$, i.e., $r=\sqrt{R_\parallel^2+z^2}$. Thereby, we find 
\begin{eqnarray} %
&& \mathcal{J}_2 = e^{-\mathrm{i}\pi/4} k_2^2 \sqrt{ \frac{ 1 }{ 2\pi k_2 r\sin\theta_1} } \int_{C_{2\eta}} d\eta \sqrt{\cos\eta}  \sin\eta \nonumber\\
&& \times R_{TM,TM}^{12}(\eta) e^{\mathrm{i} k_2 r\sin\theta_1\cos\eta + \mathrm{i}(r\cos\theta_1 + z')\sqrt{ k_1^2 - k_2^2 \cos^2\eta } }\;. \nonumber\\
\end{eqnarray}

Next we determine the saddle point of $\mathcal{J}_2$ by choosing the stationary phase as only  $\varphi\left(\eta\right) = \mathrm{i}k_2r\sin\theta_1\cos\eta + \mathrm{i}r\cos\theta_1\sqrt{ k_1^2 - k_2^2 \cos^2\eta }$, according to Ref. \cite{Banhos}. Through $\varphi^{\prime} \left(\eta_s\right)=0$ results that there are two roots: $\eta_{s}=0$ corresponding to $k_\parallel=k_2$, which we select as the saddle point, and $\tilde{\eta}_{s}=-\pi/2+\mathrm{arcsin}(k_1\sin\theta_1/k_2)$ corresponding to the saddle point employed to evaluate $\mathcal{J}_1$. This choice will lead to the full stationary phase $\mathrm{i}k_2r \sin \theta_1 - r\cos\theta_1 \sqrt{ k_2^2- k_1^2 }$.   At this stage, the steepest descent path is specified on the $\eta$-plane by demanding the next condition $\mathrm{Im} \left[ \varphi\left(\eta\right) \right] = \mathrm{Im} \left[ \varphi\left(\eta_s\right) \right]$ implying that $\mathrm{Im} [ \mathrm{i}k_2r\sin\theta_1\cos\eta +\mathrm{i}r\cos\theta_1 \sqrt{ k_1^2 - k_2^2 \cos^2\eta }\, ] =\,$$\mathrm{Im} [ \mathrm{i}k_2r\sin\theta_1 + \mathrm{i}r\cos\theta_1\sqrt{ k_1^2 - k_2^2 } \, ]$ over $C_{2\eta}$. For this integral, we do not need to shift the saddle point to the origin because it is already done. So, the next step is to introduce the conformal transformation $u^2/2=\varphi\left(0\right)-\varphi\left(\eta\right)=\mathrm{i}k_2r\sin\theta_1\left(1-\cos\eta\right) - r\cos\theta_1 ( \sqrt{ k_2^2 - k_1^2 } - \sqrt{ k_2^2\cos^2\eta - k_1^2 } \,)$ yielding to 
\begin{eqnarray}      
&& \mathcal{J}_2 = e^{-\mathrm{i}\pi/4} k_2^2 \sqrt{ \frac{ 1 }{ 2\pi k_2 r\sin\theta_1} } e^{ \mathrm{i}k_2r\sin\theta_1}  \nonumber\\ 
&& \times e^{- r\cos\theta_1\sqrt{ k_2^2 - k_1^2 } } \int_{C_{2u}} du \frac{ d\eta }{ du }F_2(u)e^{-u^2/2} \,, 
\end{eqnarray}
with 
\begin{eqnarray}     
F_2(u) & = & \cos^{1/2}[\eta(u)] \sin[\eta(u)] R_{TM,TM}^{12}[\eta(u)] \nonumber\\  
&& \times e^{ -z' \sqrt{ k_2^2 \cos^2[\eta(u)] - k_1^2 } } \;.
\end{eqnarray}
Again this integral is real and a change of variable is needed , but the latter is cumbersome to implement. To simplify the task, we will explode the fact that the dominant term of $\mathcal{J}_2$ lies in a vicinity of $u=0$ due to the presence of the Gaussian function $e^{-u^2/2}$. This will allow us to expand $F_2(u)$ and $d\eta/du$ in series around $u=0$. To this aim, we need first the expansion
\begin{eqnarray}
\frac{ u^2 }{ 2 } = \frac{ 1 }{ 2 }\left(  \mathrm{i}k_2r\sin\theta_1  - \frac{ k_2r\cos\theta_1 }{ \sqrt{ 1 - n^2_1/n_2^2 } } \right) \eta^2 + \mathcal{O}(\eta^4)\;.  \quad \;
\end{eqnarray}
Inverting this series yields to
\begin{eqnarray}
\eta &=& a_0 u + \frac{ a_2 }{ 3 }u^3 + \mathcal{O}(u^5) \;,\\
\frac{ d\eta }{ du } &=& a_0 + a_2 u^2 + \mathcal{O}(u^4) \;,
\end{eqnarray}
which immediately leads to
\begin{equation}
a_0 = \left[ \mathrm{i}k_2r\sin\theta_1  - \frac{ k_2r\cos\theta_1 }{ \sqrt{ 1 - n^2_1/n_2^2 } } \right]^{-1/2} \;.
\end{equation}
For $F_2(u)$ the required expansions at order $\mathcal{O}(u^2)$ are 
\begin{eqnarray}
\cos^{1/2}[\eta(u)] &=& 1  \;\; , \;\; \sin[\eta(u)] = a_0u  \;, \nonumber\\
R_{TM,TM}^{12}[\eta(u)]  &=& 1 + \frac{ n_2 a_0 u }{ \sqrt{ n_1^2 - n_2^2 } } \nonumber\\
&& \times \left( \frac{ \mu_1 }{ \mu_2 } - \frac{ \varepsilon_1 }{ \varepsilon_2 } + \frac{ \Delta^2 }{ \varepsilon_2 \mu_1 \mu_2^2 } \right)  . 
\end{eqnarray}
Inserting these expansions into $\mathcal{J}_2$ and carrying out the integral over $u$ finally results in
\begin{eqnarray} 
&& \mathcal{J}_2 = e^{-\mathrm{i}\pi/4} k_2^2 a_0^3 \sqrt{ \frac{ 1 }{ k_2 r\sin\theta_1} } e^{ \mathrm{i}k_2r\sin\theta_1 - r\cos\theta_1\sqrt{ k_2^2- k_1^2 } } \nonumber\\
&& \times e^{ -z' \sqrt{ k_2^2 - k_1^2 } } \left( \frac{ n_2^2 \mu_1^2 - n_1^2 \mu_2^2 + \Delta^2 }{ n_2 \mu_1 \mu_2 \sqrt{ n_1^2 - n_2^2} } \right) \;, \label{final integral J2}
\end{eqnarray}
where we identify the term in parenthesis as $\tilde{R}_{TM,TM}^{12}$ given by Eq. (\ref{Fresnel crossed 1}). After substituting the results (\ref{final integral J1}) and (\ref{final integral J2}) into Eq. (\ref{Integrals I and Js}), we complete the calculation of $\mathcal{I}_1$.

By adapting this procedure to the remaining integral $\mathcal{I}_2$ defined in Eq. (\ref{integral I2}), we obtain
\begin{equation}
\mathcal{I}_2 = \mathcal{J}_3 +H(\theta_1-\theta_1^{disc})\mathcal{J}_4\;,
\end{equation}
with $\theta_1^{disc}$ again given by Eq. (\ref{theta UH}) and 
\begin{eqnarray}
\mathcal{J}_3 &=& \frac{ e^{\mathrm{i}k_1r} }{ \mathrm{i}r } R_{TE,TM}^{12} (\theta_1)e^{ \mathrm{i}k_1z'\cos\theta_1 } \;, \\
\mathcal{J}_4 &=& \frac{ e^{-\mathrm{i}\pi/4} k_2^2 a_0^3 }{ \sqrt{ k_1^2 - k_2^2} } \sqrt{ \frac{ 1 }{ k_2 r\sin\theta_1} } e^{ \mathrm{i}k_2r\sin\theta_1  - r\cos\theta_1\sqrt{ k_2^2- k_1^2 } } \nonumber\\
&& \times e^{ -z' \sqrt{ k_2^2 - k_1^2 } } \left( - \frac{ 2n_1 \Delta }{ n_2 \mu_1 \sqrt{ n_1^2 - n_2^2} } \right) \;, 
\end{eqnarray}
where the term in parenthesis is identified as $\tilde{R}_{TE,TM}^{12}$ of Eq. (\ref{Fresnel crossed 2}).

Finally, by substituting the final forms of both integrals $\mathcal{I}_1$ and $\mathcal{I}_2$ into Eq. (\ref{Ex1 int}) and after carrying on the remaining partial derivatives, one arrives at the $x$-component of Eq. (\ref{E reflex}).


\end{document}